\documentclass[fleqn,usenatbib]{mnras}

\usepackage{newtxtext,newtxmath}

\usepackage[T1]{fontenc}
\usepackage{ae,aecompl}

\usepackage{graphicx}	
\usepackage{amsmath}	
\usepackage{amssymb}	
\usepackage{todonotes}
\usepackage{mathrsfs}
\usepackage{longtable}

\title[PTF and iPTF Type Ia SN R-band light-curves]{R-band light-curve properties of Type Ia supernovae from the (intermediate) Palomar Transient Factory}

\author[Papadogiannakis S. et al.]{S. Papadogiannakis$^{1,2}$\thanks{E-mail: semeli@fysik.su.se},
	A. Goobar$^{1,2}$,
    R. Amanullah$^{1,2}$,
    M. Bulla$^{1,2}$,
    \newauthor
    S. Dhawan$^{1,2}$,
    G. Doran$^{3}$
    U. Feindt$^{1,2}$,
    R. Ferretti$^{1,2}$,
    L. Hangard$^{1,2}$,
    D. A. Howell$^{4,5}$,
    \newauthor
    J. Johansson$^{6}$,
    M. M. Kasliwal$^{7}$, 
    R. Laher$^{8}$,
    F. Masci$^{8}$,
    A. Nyholm$^{9}$,
    E. Ofek$^{10}$, 
    \newauthor
    J. Sollerman$^{9}$
	and L. Yan$^{6}$
	\\
    \\
	$^{1}$Department of Physics, Stockholm University, SE 106 91 Stockholm,  Sweden\\
	$^{2}$Oskar Klein Centre, Department of physics, Stockholm University, SE 106 91 Stockholm,  Sweden\\
    $^{^3}$Jet Propulsion Laboratory, California Institute of Technology, USA
	$^{4}$Las Cumbres Observatory, University of California, Santa Barbara, USA \\
    $^{5}$University of California, Santa Barbara, Department of Physics, Broida Hall, Santa Barbara, CA, USA 93106 \\
    $^{6}$Department of Physics and Astronomy, Division of Astronomy and Space Physics, Uppsala University, Box 516, SE 751 20 Uppsala, Sweden \\
    $^{7}$ Caltech Optical Observatories, California Institute of Technology, Pasadena, CA 91125, USA \\
      $^{8}$ Infrared Processing and Analysis Center, California Institute of Technology, Pasadena, CA,
91125, USA \\
$^{9}$ Department of Astronomy and The Oskar Klein Centre, Stockholm University, SE-106 91 Stockholm, Sweden \\
     $^{10}$ Benoziyo Center for Astrophysics, Weizmann Institute of Science, 76100 Rehovot, Israel
}

\date{Accepted XXX. Received YYY; in original form ZZZ}

\pubyear{2018}

\begin{document}
	\label{firstpage}
	\pagerange{\pageref{firstpage}--\pageref{lastpage}}
	\maketitle
	
	\begin{abstract}
We present the best 265 sampled R-band light curves of spectroscopically identified Type Ia supernovae (SNe) from the Palomar Transient Factory (PTF; 2009-2012) survey and the intermediate Palomar Transient Factory (iPTF; 2013-2017). A model-independent light curve template is built from our data-set with the purpose to investigate average properties and diversity in our sample. We searched for multiple populations in the light curve  properties using machine learning tools. We also utilised the long history of our light curves, up to 4000 days, to exclude any significant pre- or post- supernova flares. From the shapes of light curves we found the average rise time in the R band to be $16.8^{+0.5}_{-0.6}$ days.
Although PTF/iPTF were single-band surveys, by modelling the residuals of the SNe in the Hubble-Lema\^{i}tre diagram, we estimate the average colour excess of our sample to be $<$E$($B$-$V$)> \approx 0.05(2)$ mag and thus the mean corrected peak brightness to be $M_R = -19.02\pm0.02$ $+5 \log( {\rm H}_0 [{\rm km} \cdot{\rm s}^{-1} {\rm Mpc}^{-1}]/70)$ mag with only weakly dependent on light curve shape. The intrinsic scatter is found to be $\sigma_R = 0.186 \pm 0.033$ mag for the redshift range $0.05<z<0.1$, without colour corrections of individual SNe. Our analysis shows that Malmquist bias becomes very significant at z=0.13. A similar limitation is expected for the ongoing Zwicky Transient Facility (ZTF) survey using the same telescope, but new camera expressly designed for ZTF.  
	\end{abstract}
	
	
	
	
	\section{Introduction}
    Type Ia supernovae (SNe) are understood by now to be thermonuclear explosions of white dwarfs. However, the mechanism of the explosion remains unknown. The leading theories involve binary interaction  with two different scenarios; the single degenerate (SD) and the double degenerate (DD) scenario involving a giant or main sequence companion star or a white dwarf companion, respectively
    \citep[see][for a recent review]{2016IJMPD..2530024M}.
	Despite the lack of theoretical certainty about progenitors, type Ia SNe have proven very useful in cosmology as ``standardisable" distance estimators, which led to the discovery of the accelerating expansion of the universe \citep{Riess98,Perlmutter99a} attributed to the existence of a new cosmic constituent dubbed ``dark energy" \citep[see][for a review]{2011ARNPS..61..251G}.
	
	Following the discovery of dark energy, many studies have focused on increasing the precision and accuracy of the cosmological parameters derived from type Ia SNe combined with other cosmological probes \citep[e.g.][]{2014A&A...568A..22B, 2017arXiv171000845S}. Both statistical and systematic uncertainties need to be improved to discern between dark energy models, see e.g. \citet{2017JCAP...07..040D}. The systematics include, but are not limited to possible brightness evolution over cosmic time, cross-calibration of different instrument, telescope data and properly accounting for extinction by dust in the line of sight. One way to study the systematic uncertainties is to investigate large samples of nearby and distant SNe, as shown in many works in the literature, e.g. by the SDSS-II and SNLS collaborations \citep{2009ApJS..185...32K, 2011ApJ...737..102S, 2014A&A...568A..22B}. Other important contributions include results from  PTF \citep{Maguire14}, the Supernova Cosmology Project \citep[SCP,][]{2010ApJ...716..712A,  2012ApJ...746...85S} and from PanSTARRS1 \citep{2014ApJ...795...44R}. Another approach to better understand systematics is to study nearby individual SNe to probe the SN physics. Examples of such studies based on Palomar Transient Factory (PTF) and its successor, the intermediate Palomar Transient Factory (iPTF) include  \citet{2011Natur.480..344N}, \citet{2012Sci...337..942D}, \citet{2014ApJ...784L..12G}, \citet{2015ApJ...799..106G}, \citet{2015Natur.521..328C}, \citet{2015A&A...578A...9H} and \citet{2017ApJ...848...59M}. 
    
    In this paper we use a large homogeneous data set of low-redshift SNe Ia in a single photometric band from the Palomar 48-inch Oschin Schmidt Telescope to address some of the uncertainties associated with their use in cosmology. PTF and iPTF were two surveys dedicated to finding, among other things, SNe within days from explosion \citep{2009PASP..121.1334R}. The survey imaged hundreds of square degrees of the sky, twice or more times per night. This enabled us to build  light curves of the transients, i.e., follow their brightness over time. Through this strategy two different time scales were probed simultaneously: a longer one over the years the survey ran and a shorter intra-night timescale. The large field of view of the PTF/iPTF, 7.26 deg$^2$, allowed us to cover a large part of the sky and thus building a statistical sample of type Ia supernovae detected in a similar fashion, and minimizing selection effects.
    
	We present observations in the R band for the SNe with the most complete coverage. These are used to explore the light curve properties and possible signs of yet unknown diversity among SNe Ia. 
     For the light curve as a whole, we use a non-parametric fitting method, Gaussian processes, to generate a smooth version of the light curves in order to look for signs of multiple SN Ia populations and to study intrinsic dispersion at different epochs (see Section \ref{Section:template}). In the same Section, we also use the light curves in 3 different redshift bins to look for diversity in a given epoch at different cosmic times.
 We present average photometric properties of the sample, e.g., the rise-time distribution light curve (Section \ref{Section:lcfitting}), and the dispersion of the light curves at various epochs (Section \ref{Section:GPtemplate}).  We utilise the long history of detections before and after the supernova light is visible to set limits on a pre- and post-explosion event in Section \ref{Section:prelims}. From the distribution of residuals in the Hubble-Lema\^{i}tre diagram, we explore if there is a correlation with light curve shape in the R band (Section \ref{Section:hubble}) and the stellar mass of the host galaxy (Section \ref{Section:mass}). Furthermore, we estimate the mean free path due to scattering by dust along the line of sight, even without colour information.
 
	In a follow-up paper we will present the spectra used to classify the SNe and determine the redshift of the SNe in this study, as well as detailed a analysis  of their spectroscopic properties, and use machine learning techniques to relate these to the photometric properties shown in this work. 
    
\section{The data set}\label{Section:dataset} 
	\subsection{The PTF and iPTF transient surveys}
	PTF and iPTF surveyed the sky regularly to discover new transients with an unprecedented large field of view.
     The survey was conducted in a single filter at a time, mostly in the Mould R band (wavelength range $5800$-$7300$ \AA), but data in g band (wavelength range $3900$-$5600$ \AA) were also collected during some periods. Narrow $H_{\alpha}$ filters at several recession velocities were used during the 2-5 days closest to the full moon each month. The magnitude limit of the survey was 20.5 and 21 magnitudes for R and g band respectively in the PTF system. In this paper, we focus on the R-band observations.

	PTF and iPTF performed a non-targeted survey by imaging the sky 1-5 times per night with exposures on the same field (at least 40 minutes apart) and then performing difference imaging, in order to discover new transients. 50\% of the observations are taken with a 1 day cadence or shorter and 70\% within 4 day cadence excluding the intra-night cadence which is the most common (43 or 63 minutes apart). 
    The reference images were taken in 2009 and 2012 for PTF and iPTF, respectively, for the majority of the fields. A non-targeted survey means that no particular part of the sky was imaged in the survey, thus minimising the bias associated with targeted searches, e.g. finding transients only in well-resolved host galaxies\footnote{Note that iPTF was not completely blind as it followed a Census of the Local Universe catalogue of galaxies within 200 Mpc (Cook et al. in prep) for 8 months during the spring and autumn of 2013.}. In addition, since we use data only from a single instrument and photometric band, other systematic effects are minimised. This makes PTF and iPTF ideal for minimising the sampling bias.

After running through an image-subtraction pipeline the measured parameters from the extracted sources were analysed using a machine learning algorithm \citep{2008AN....329..284B}. This algorithm sets a score on the likelihood that each candidate is an astrophysical transient, which is used to discard the many false candidates that are found by the pipeline. For the PTF collaboration, this was done in a combination of ``Supernova zoo participants" \citep{2011MNRAS.412.1309S} to train the algorithm and an effort of the collaboration where the top candidates were screened by team members and sent for spectroscopic follow-up. The overall supernova detection performance of the PTF survey is explored in \citet{2017ApJS..230....4F} and the iPTF survey efficiency estimation is work in progress. For the iPTF data the top candidates were selected solely by people from the collaboration.
		
	This survey strategy and rapid follow-up enabled discoveries of transients close to the last non-detection limits. The mean of the first detection point in time for our SNe is -12 days, compared to -4 days in the low redshift sample presented by \citet{2014A&A...568A..22B}. A histogram of the first detection points of both surveys is shown in Figure \ref{fig:first_detection} in Section \ref{Section:hubble}, where the implications are also discussed.
    	
	\subsection{Photometry and Calibration} \label{Section:photcalib}
	All photometric data used in this paper are in the Mould R band \citep[see][and Appendix \ref{Section:filter}]{2009PASP..121.1395L}, corrected for quantum efficiency of the instrument. The PTF image processing is described in \cite{2014PASP..126..674L}.
	We used the PTF-IPAC forced photometry pipeline by \citet{2017PASP..129a4002M}, to produce the light curves. The procedure to process the PTF-IPAC pipeline photometry in light curves used in our analysis is described in detail in Appendix \ref{appendix:forced}. 
    
	The photometric pipeline performs difference imaging on a fixed position, in this case, the position of the supernova as determined at discovery, to remove the host galaxy contamination. A point spread function (PSF) fit is then performed at this position for each of the images. Where calibration against images from the Sloan Digital Sky Survey (SDSS) was not possible, a field observed during the same night was used.
	
	The error estimates of each data point take into account the goodness of fit of the PSF, the overall zero point at the time of observation as compared to SDSS wherever possible in order to get the absolute photometry. Note that the magnitudes used in this paper are magnitudes are in the PTF system \citep[rather than the AB system, see conversion formulae in][]{2012PASP..124...62O}, and thus have not been corrected for the colour of SNe Ia. The repeatability between different CCD chips for the same stars is better than 0.03 mag in 95\% of cases, see \citet{2012PASP..124...62O}. There are additional systematics that were deemed sub-dominant, including incorrect PSF template estimation, uncertainties in the SN position and astrometric calibration which determine the central position of the PSF fit. 
    
	\subsection{The type Ia SNe sample}
	In this paper we examine the statistical properties of 265 out of 2059 spectroscopically confirmed type Ia supernovae from PTF and iPTF (from 2009-2017), selected due to their well sampled  R-band light curves (see criteria in Section \ref{Section:template}). We do not exclude any SNe based on their spectroscopic sub-classification. Due to the observing strategies in 2015 and 2017 no SN Ia was included from these years.
    
    We classify the supernovae using Supernova Identification software SNID \citep{blondin07} using the version 2.0 templates. We select the 5 best fits that pass the SNID criteria ``good" and choose the most common type from these. We then visually inspect the best fits to be certain of the typing.

For 169 of the SNe in our sample, the redshift is measured from host galaxy lines in the SN spectra or from the host spectrum. When this is not available we use the SDSS spectral redshift  (15 SNe) of the host galaxy or host redshifts from NED (3 SNe) and if that is not available the median redshift of the 5 best estimates from SNID is used (56 SNe). We note that to have a precise redshift the hosts would have to be revisited to get a more accurate redshift.
	
\begin{figure*}
		\centering
		\includegraphics[width=\hsize]{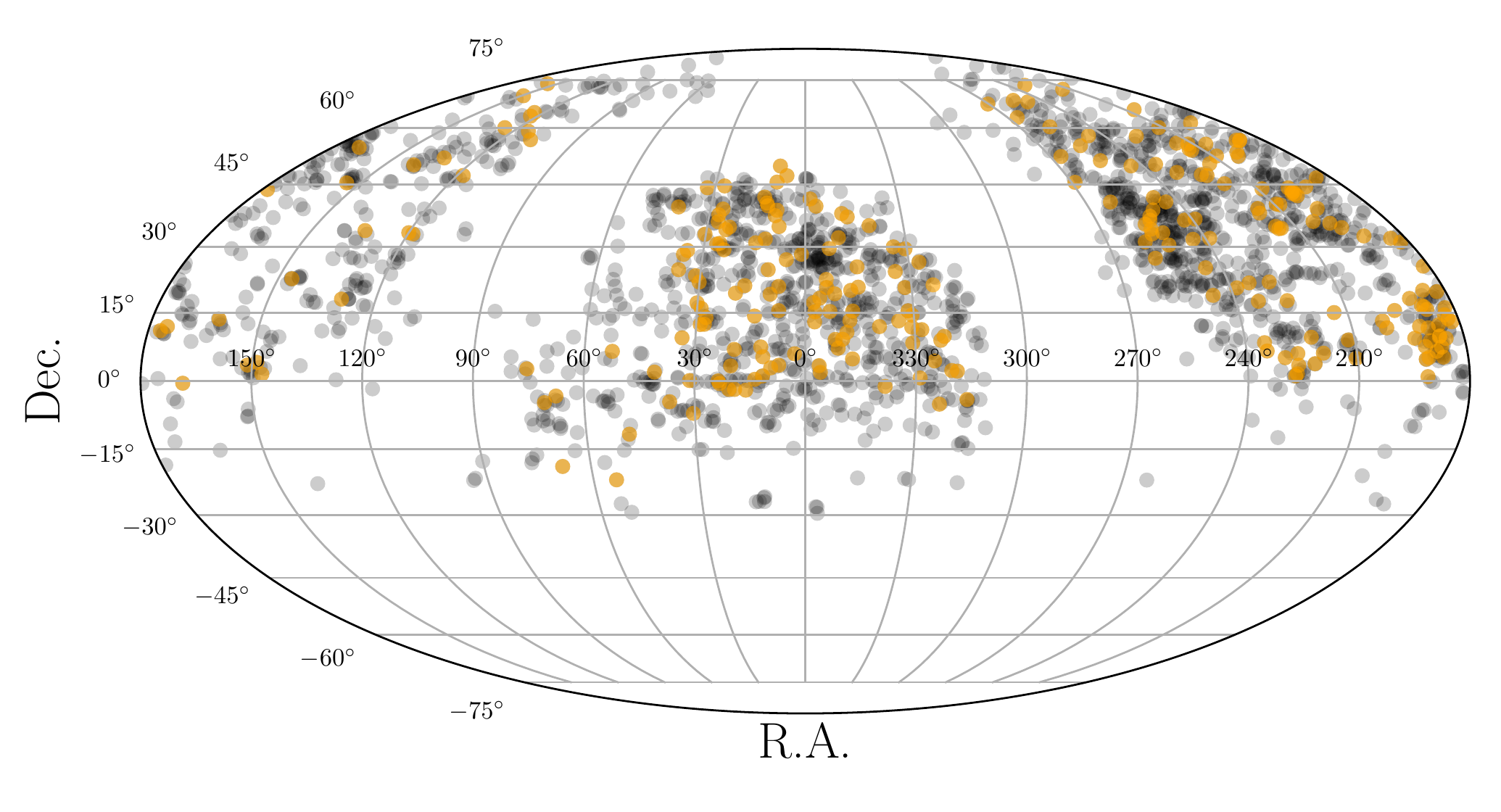}
		\caption{Right-ascension (RA) and declination (Dec) distribution of the type Ia supernovae from the PTF and iPTF surveys. In yellow points we see the 265 best sampled SNe used in this work, the black points show the rest of the type Ia SNe from the PTF and iPTF surveys. The empty regions is the location of our Milky Way galaxy and the southern hemisphere.}
		\label{fig:radec_distribution}
	\end{figure*}

	In Figure \ref{fig:radec_distribution} we show the spatial distribution on the sky of the data sample. Due to weather constraints a larger portion of well-sampled SNe are from the spring/summer half of the year. The gap in data on the northern hemisphere is from the galactic plane which obscures extragalactic SNe. The area around the galactic plane is also very crowded, i.e. filled with many stars, and thus harder to perform accurate image subtractions to find transients. 
	
	\begin{figure}
		\centering
		\includegraphics[width=9cm]{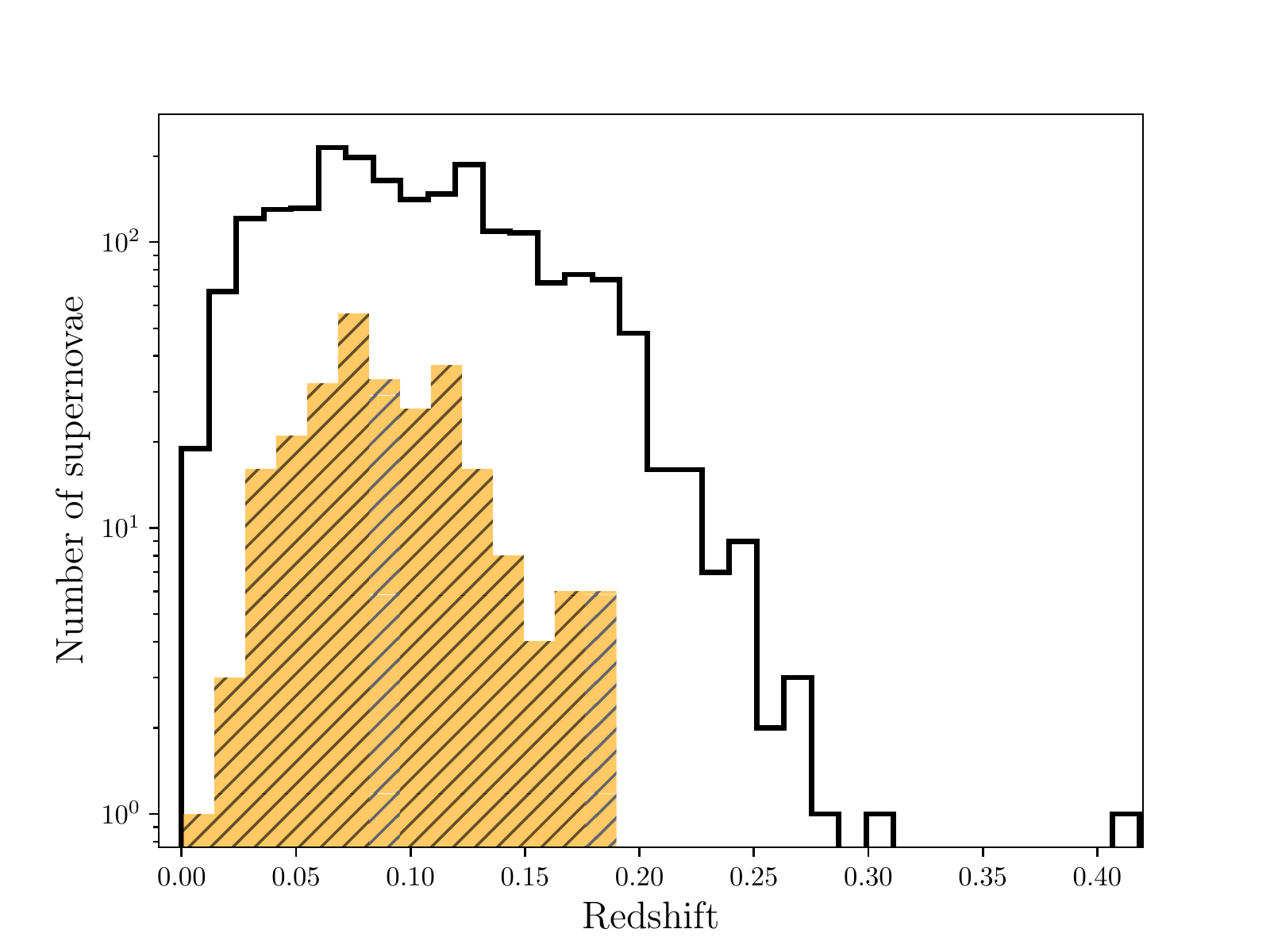}
		\caption{Redshift distribution of the PTF and iPTF SNe Ia sample. In the shaded region we show the distribution for the most well-sampled SN Ia used in this work. Note iPTF16geu, at redshift 0.4 as a significant outlier \protect \citep{2017Sci...356..291G}.}
		\label{fig:z_distribution}
	\end{figure}
	
	In Figure \ref{fig:z_distribution} we show the redshift distribution of our data sample in shaded and in comparison to the entire PTF and iPTF sample of type Ia SNe. 
	
\section{light curves and building a template}\label{Section:template}

The norm in modern cosmology with type Ia SNe is to fit a time-evolving spectral energy distribution (SED) to the light curves to extract parameters used to derive their distance,e.g. MLCS2k2 \citep{2007ApJ...659..122J}, BayeSN \citep{2011ApJ...731..120M}, SALT2 \citep{2007A&A...466...11G} , SIFTO \citep{2008ApJ...681..482C} and SNooPy \citep{2011AJ....141...19B}. In order to use our data instead of a parametrized template to fit our SNe, we here use a model that does not impose a pre-defined form to construct an empirical model template. The template is used to extract parameters such as peak magnitude and stretch, but also to study the intrinsic dispersion at different epochs along the light curves. This method, Gaussian processes, has been used for type Ia SN cosmology previously \citep[ in e.g.][]{2010PhRvL.105x1302H, Kim13, 2013PhRvD..87b3520S, 2016ApJ...832...86C} but not for large samples, mainly due to its computationally intensive nature. We start by aligning the light curves in Section \ref{Section:align} and then perform Gaussian processes in Section \ref{Section:GPtemplate} to obtain a template and study the light curve parameters. Throughout this paper we use the code packages \texttt{Astropy} version 2.0.4 \citep{2018arXiv180102634T}, \texttt{Matplotlib} \citep{Hunter:2007}, \texttt{Scipy} \citep{scipy} version 1.0.0, \texttt{numpy} version 1.14.1 and \texttt{sncosmo} version 1.5.3 \citep{barbary_2014_11938} for our data analysis.
	
	\subsection{Quality cuts and aligning the light curves}\label{Section:align}
	We align the light curves in time and normalise their magnitudes, such that zero is the peak magnitude. 
	
	The following conditions have been set for the supernovae included in the sample:
	
	\begin{itemize}
		\item[1.] More than 10 data points in the light curve, at least 3 before and 5 after time of peak.
		\item[2.] At least 4 points within $\pm 5$ days of the peak.
		\item[3.] Data spanning at least 15 days.
		\item[4.] Not located in a known quasar or active galactic nucleus (AGN).
	\end{itemize}
    
	From the 2059 spectroscopically confirmed SNe Ia in the survey we had 1705 in the R-band from these we apply the first cut with data from the Nugent photometric pipeline (an aperture photometry pipeline) that was the real-time pipeline used in the surveys and the remaining cuts with the PSF based PTF-IPAC pipeline. 1104, 133, 70 and 7 SNe are cut by the first, second, third and fourth condition respectively. The reason for having such strict constraints is to ensure an accurate template and be well-sampled enough to probe the different science questions investigated further in the paper, such as early light-curves. In future work less strict cuts can be made for different science cases. The first, second and third conditions are there to pinpoint the peak and the fourth to eliminate high intrinsic noise in supernovae light curves caused by their environment. The last condition only accounts for registered AGN activity in the host galaxy. For the case of SN\,2014J (or iPTF14jj) we exclude this from our light curve template analysis due to saturated data point, however we include it in Section \ref{Section:prelims} since that part of the light curve is unaffected by the saturated points.
	
	First, the peak of the light curve was estimated by using the brightest point in the light curve and then fitting the interpolation of a well sampled supernova from our sample, PTF10hmv, and selecting the peak that minimises $\chi^2$. We then check that the conditions are fulfilled and correct the remaining light curves for cosmological time-dilation and align in them in time and magnitude according to this initial peak estimate. 
	
	From this initial alignment we now K-correct the light curves, apply our cuts and minimise the modified ${\chi^2}$, 
	\begin{equation}
		\mathcal{Q}^2 = \left.{\displaystyle\sum_{i}^{N} \left( \frac{m_i - m_T(d_i + \delta t) + A}{\sigma_{phot, i}} \right)^2} \middle/ \right. N^4, 
	\end{equation}
over the parameters time $\delta t$, and magnitude normalisation $A$. $m_T(t)$ is the magnitude of the template at time $t$, $(d_i, m_i)$ are the normalised times and magnitudes and $\sigma_{phot, i}$ is the photometric error.

Since only the points between -20 and +100 days with respect to maximum light contribute to the $\chi^2$, we can trivially obtain a perfect fit by shifting the points until only one is left in range. To counteract this, we need to encourage the loss function to include points. One possible way is to include some penalty for bright points outside of the range, but this would not be effective since there are some photometric artefacts. Instead, we decided to explicitly reward the inclusion of points by dividing $\chi^2$ by $N^2$. Several other factors were tried (such as $N$, $N^3$, $\sqrt(N)$...), but $N^2$ yielded the most well-aligned light-curves. Higher factors, like $N^3$, compress the light-curve in order to add more points, while lower factors like $N$ and $\sqrt(N)$ suffer similar problems to a normal $\chi2$. This has the consequence of adding a bias on the stretch factor of the SNe which we avoid by using the $N^2$ factor. This initial template is made with data from -20 days to +80 days since this is the interval for which we have K-corrections and sufficient data.
	The K-correction \citep{Oke68, 1996PASP..108..190K} takes the observed magnitude and converts into the magnitude it would have had in a common rest-frame which requires the SED of a supernova. We used the SED of \citet{Hsiao07} consisting of about 600 spectra in the time span of -20 to +85 days with respect to B-band maximum, and adapted equation 2 in \citet{Oke68} for K-corrections in the same band, in this case the P48 R-band, $K_R$. Here $F(\lambda)$, $S_R(\lambda)$ and $z$ are the spectral energy distribution for a given wavelength $\lambda$, the filter transmission for the same wavelength and redshift respectively.  
	
	\begin{equation}
		K_R = 2.5 \log_{10}(1+z) + 2.5 \log_{10}\left( \frac{\int F(\lambda)S_R(\lambda)d\lambda}{\int F\left(\frac{\lambda}{(1+z)}\right) S_R(\lambda)d\lambda}\right) 
	\end{equation}
	
	The K-correction in R-band evolves with epoch and vary between -0.01 and -0.35 magnitudes (for z=0.2). For the entire PTF and iPTF samples the mean K-correction is -0.25 magnitudes. Uncertainty in K-corrections is expected to be larger for peculiar supernovae since the template is made with ``normal" type Ia supernovae. We estimate the error in our K-corrections by comparing our fits to SALT2 fits. 
    
 We fit the SALT2 model to the (i)PTF r-band lightcurves using \texttt{sncosmo}. Since we were only using data in a single band, we fixed the color parameter $c$ to 0 but applied observer-frame extinction based on Milky Way dust. Most lightcurves contain limits from observations of their location that were made years before and/or after the SN exploded. Since we do not gain much for the SALT2 fit from most of those limits, we discarded any data 30 days before the first data point with $S/N > 5$ and 30 days after the last point with that significance. Based on the best-fit values for the remaining parameters we then calculated the rest-frame peak brightness in r-band (as well as the standard B-band). When calculating the the $\chi^2$-values listed in Table \ref{table:salt2}, we excluded the points that fall outside the definition range of the SALT2 model that was fit (and which otherwise would lead to very low values of $\chi^2$/d.o.f.\ because the limits will perfectly match the model flux, which is set to zero outside the definition range). We then use these fits to estimate the K-correction error by fitting a Gaussian to the difference between the maximum magnitude from the SALT2 fits and our fits to get the variance between the two, which is found to be 0.046 mag. This is a conservative estimate, as other sources of error cannot be excluded.
    
	When this first fitting has been done, we make sure that the conditions are still fulfilled, and then proceed to doing a second fit. This time another free parameter is allowed, measuring the light curve width, stretch $S$. Stretch is defined to be a multiplicative factor that measures the width of the light curves, thus $S<1$ implies a narrow shape, $S>1$ a broad shape and $S=1$ a shape that exactly matches that of the template similar to what was done in \citet{1997ApJ...483..565P}. The time $t$ in days is thus defined to be, 
	\begin{equation}
		t = t_0 \times S.
	\end{equation}
	
	The light curves are fitted to the template created from the first fit minimising
	\begin{equation}
		\tilde{Q} ^2 = \left.{\displaystyle\sum_{i}^{N} \left( \frac{m_i- m_T(d_i \times S + \delta t) + A}{\sigma_{phot, i}} \right)^2} \middle/ \right. {N^4},
	\end{equation}
	over the parameters $A$, $\delta t$ and $S$.  
	
\begin{figure*}
		\centering
		\includegraphics[width=\hsize]{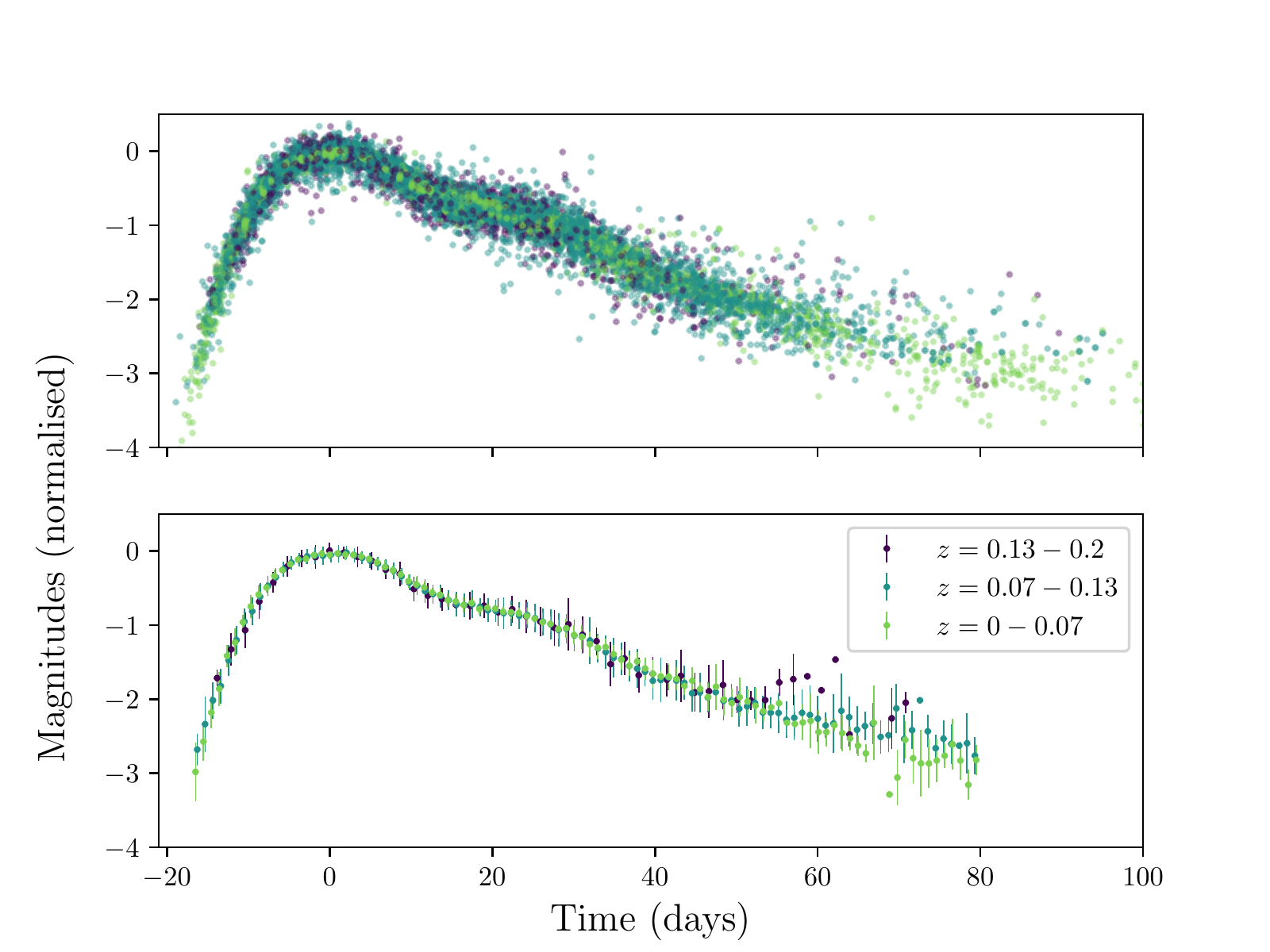}
		\caption{All aligned light curves. The data points are colour-coded according to redshift as shown in the legend. In the upper panel, we show the un-binned data and in the lower panel we show binned data points, in 3 redshift ranges.}
		\label{fig:lc_aligned}
	\end{figure*}

	As shown in the upper panel of Figure \ref{fig:lc_aligned} we see the final 265 aligned  and averaged SNe and in the lower panel of Figure \ref{fig:lc_aligned} the same but binned in 3 redshift ranges.
	
	From the starting sample of 2059 supernovae, 265 remained at the end for the R-band after quality cuts were applied. Table \ref{table:process} shows at what stage the supernovae drop out. The first step selects the R-band light curves with the initial maximum estimate of maximum light to fulfil the conditions.

	\begin{table}
		\caption{Total number of SNe in the sample after each respective process in preparing the light curves for the template.}              
		\label{table:process}      
		\centering                                      
		\begin{tabular}{c c l l}          
			\hline\hline                        
			Process & Number of SNe \\    
			\hline                                   
			Conditions met using initial maximum & 391 \\      
			K-corrections \& fitting of maximum & 344  \\
			Stretch correction added & 265 \\
			\hline                                             
		\end{tabular}
	\end{table}
	
	We correct for Milky way extinction at the position of the supernova using the maps of \citet{2018MNRAS.478..651G}, implemented in the package \texttt{dustmaps}\footnote{https://github.com/gregreen/dustmaps}to get E(B-V), i.e. the colour excess. We then use, 
	\begin{equation}
		A_R = \frac{A_V}{R_V} \frac{\lambda_B}{\lambda_R} \left(\frac{\lambda_V -\lambda_R}{\lambda_V-\lambda_B} \right) + A_V
	\end{equation}\label{eq:A_r}
	to find the extinction in the R-band, $A_R$ due to Milky way extinction. We assumed the total-to-selective extinction parameter, $R_V = 3.1$. Here $\lambda_i$ is the central wavelength in the $i^{th}$ band and $A_V$ is the extinction in the V-band. The average is found to be $0.095$ magnitudes in the R-band. 
    
    We do not set an upper limit requirement on $A_V$ in our sample, hence the largest galactic E(B-V) among our SNe is 0.79 mag compared to the 0.15 mag limit set by \citet{2014A&A...568A..22B} for inclusion in the Hubble-Lema\^{i}tre diagram.

After these corrections the last step performs in addition a stretch correction and refits for the peak magnitude. At all processes the conditions to be fulfilled are rechecked. We find the root-mean-squared, rms of the aligned light curves (for all epochs) to be $0.19$ magnitudes within 5 days of the peak. The result of the aligned light curves are shown in Figure \ref{fig:lc_aligned}.
\\
\\
\\
\subsection{Gaussian Processes template}\label{Section:GPtemplate}
	
In order to get a predictive light curve template we have used Gaussian processes (GP). This method allows a non-parametric way to estimate, based on the training data (our dataset), what the predicted behaviour will be for a supernova and in addition allows deviation from this to be quantified. This method has been applied to supernova cosmology before by \citet{Shafieloo13}, \citet{Holsclaw_2010} and for modelling type Ia supernovae in \citet{Kim13}. Since Gaussian processes decay to zero outside of the data range, we perform the fitting in flux space.
\begin{figure*}
		\centering
		\includegraphics[width=17cm]{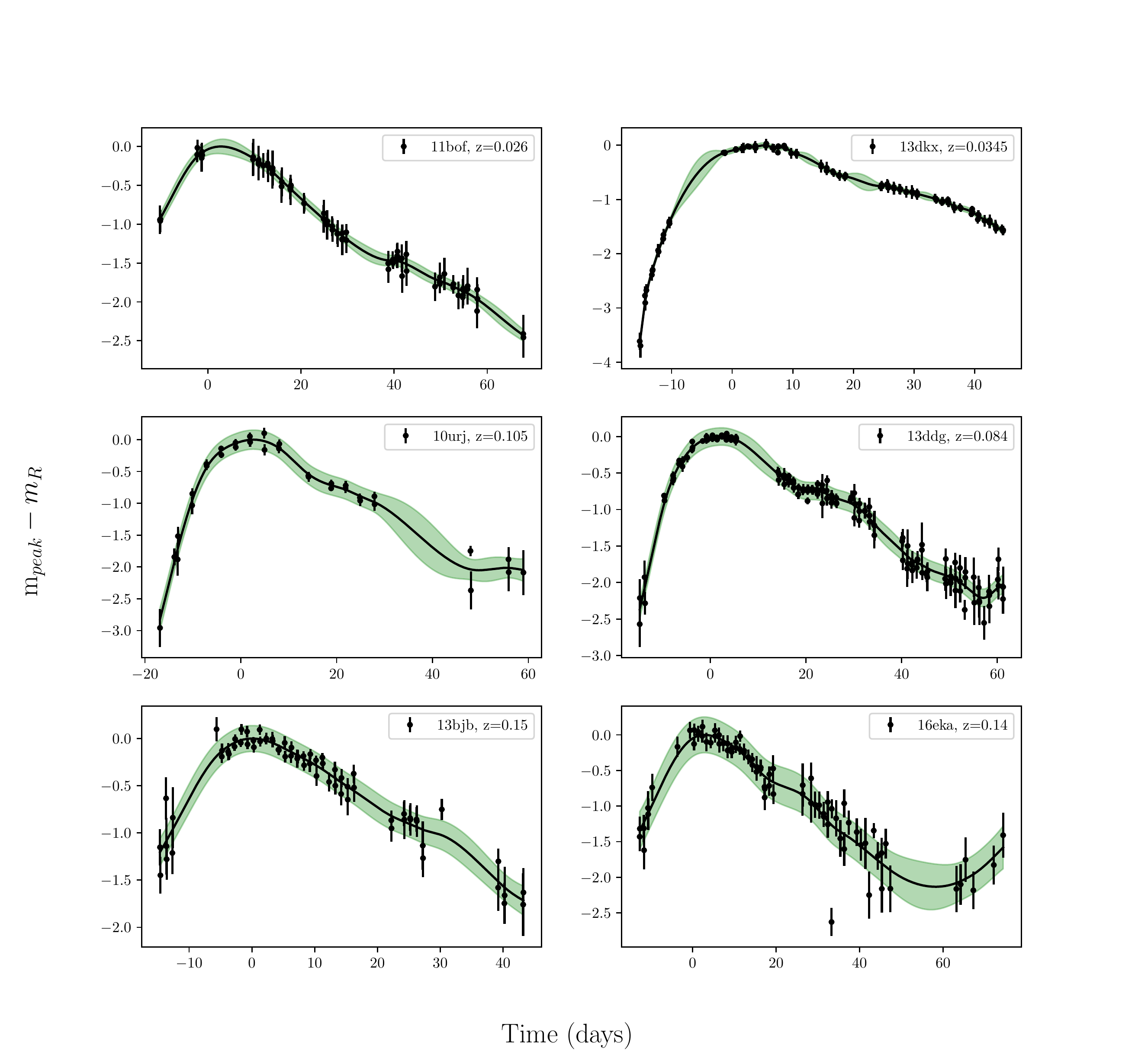}
		\caption{Six example light curves with their Gaussian processes fit with normalised apparent magnitudes, $m{peak}-m_R$ against time in days. The upper, middle and lower panel show two SNe each from the redshift bins $0-0.07$, $0.07-0.13$ and $0.13-0.2$ respectively. The shaded region shows the 1 $\sigma$ interval, as predicted by GP around the latent function shown in a solid line.} 
		\label{fig:SNexamples}
	\end{figure*}

\begin{figure*}
		\centering
		\includegraphics[width=17cm]{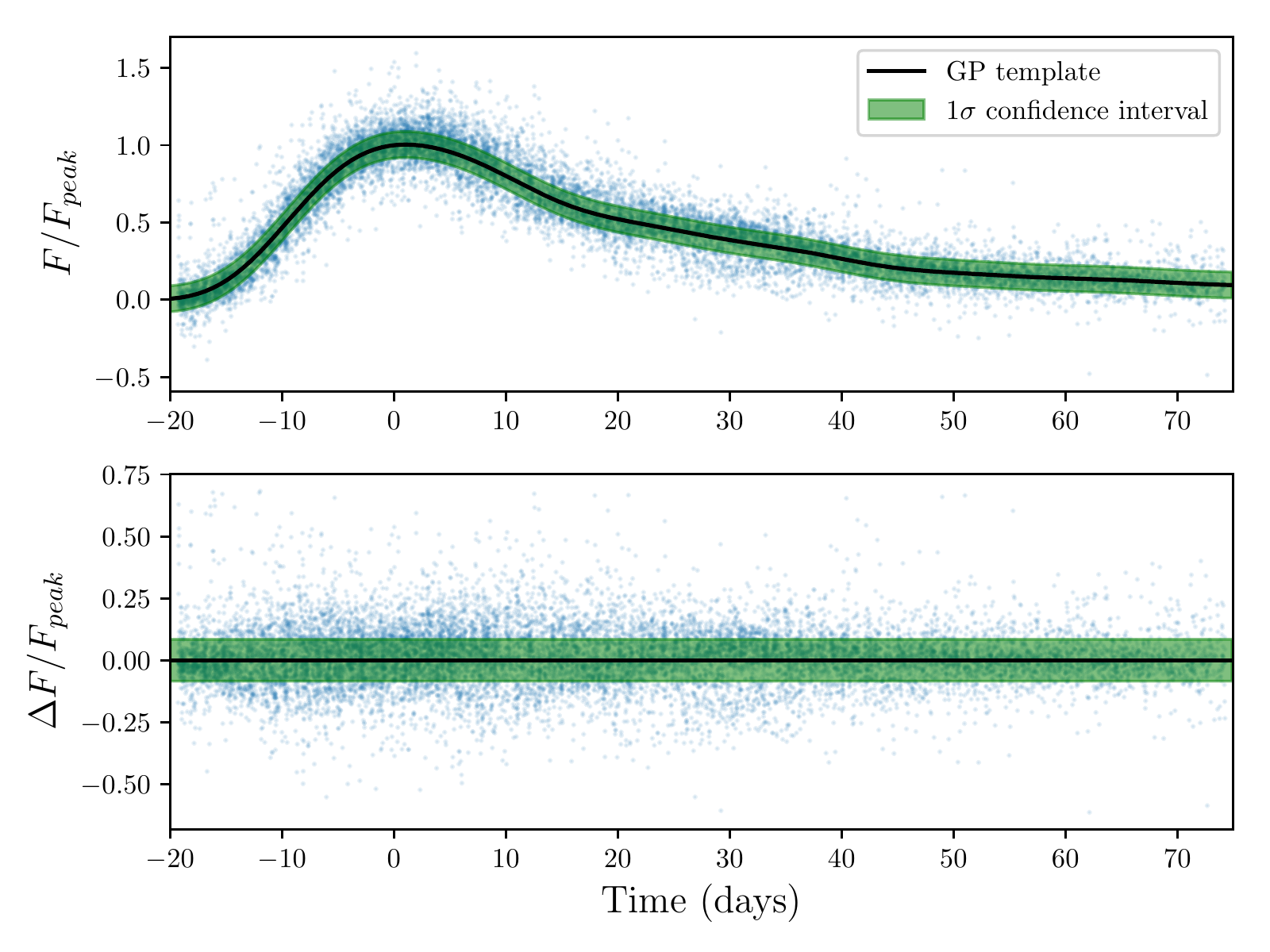}
		\caption{GP template of the combined light curves of 265 PTF and iPTF SNe Ia in flux space. The solid line shows the most likely function and the shaded region shows the 1 $\sigma$ interval, as predicted by GP. The axes show the normalised fluxes, $F/F_{peak}$, vs time in days. The lower panel shows the residuals, $\Delta F/F_{peak}$, of the template.} 
		\label{fig:template_res}
	\end{figure*}

We used heteroscedastic (accounting for the error of each data point) Gaussian processes to get a template of our light curve data sample in the R-band spanning from -20 to + 75 days with respect to maximum light. In Figure \ref{fig:SNexamples} we show what the GP fit looks like for six representative SNe in our sample, two from each of the redshift bins $0-0.07$, $0.07-0.13$ and $0.13-0.2$ respectively. The result of the template, when applied to the aligned light curves, is shown in Figure \ref{fig:template_res} with the residuals on the lower panel of the same plot and found in Table \ref{table:template}. Due to the computationally expensive nature of heteroscedastic Gaussian processes, including inverting a large matrix, the code was run on a computer cluster using 2TB of RAM. The matrix is square with the size of the number of data points, i.e. $11960 \times 11960$.  For more details on Gaussian Processes and how it was applied here see appendix \ref{appendix:GP}.

\subsubsection*{Reliability of the template}
We test the robustness of our GP template by using Monte-Carlo simulations of the light curves with random Gaussian noise proportional to the measurement error and then repeating this for light curves with the same error and a systematic offset. To get an estimate on how sensitive all the parameters, such as stretch, time of maximum and maximum magnitude, are for noise we assume that our GP template is the ``truth" and then re-fitting the simulated light curves (with added Gaussian noise proportional to the measurement error). We found that our template is robust (i.e. the standard deviation of the stretch was 0.04 for the 10,000 simulations) and use our results of the later simulation as an estimate for the error in the light curve parameters.

\subsection{Searching for multiple populations}\label{Section:populations_stretch}
We can thus trust the template and are able to examine the residuals in order to search for multiple populations. If such were found it would point to diversity in the SNe physics. To measure the intrinsic scatter around each epoch, we divide the template into time bins of 9 days and fit Gaussian Mixture models from scikit-learn version 0.19.1 \citep{scikit-learn} to each bin. The aim was to see if one Gaussian or more explain the distribution of each epoch bin better. 

To evaluate the significance of this result we used the Bayesian information criteria (BIC) from \citet{1978AnSta...6..461S}, defined in equation \ref{eq:bic}, where $N$ is the number of data points in the fit, $\mathcal{L}$ is the maximum likelihood and $k$ is the number of parameters in the model. 
    
    \begin{equation}\label{eq:bic}
	BIC \equiv - 2 \ln \mathcal{L} + k \ln N
	\end{equation}
    
    As discussed in \citet{2004MNRAS.351L..49L}, BIC tends to favour models with fewer parameters compared to the commonly used Akaike information criteria (AIC), which is why we choose BIC for the purpose of determining if there is more than one population in the supernova parameters such as stretch. The best model is the one with the lowest value of BIC and if the difference between values of BIC, $\Delta$BIC is larger than 6 it is considered that the model is favoured significantly \citep[see e.g.][]{2009ApJ...703.1374S}. Since we prefer to be conservative in declaring a potential multiple population detection we require, in addition to $\Delta BIC > 6$, that the mean of the two distributions is at least 3 $\sigma$ from each other.
    
We find that all bins are significantly better fitted ($\Delta BIC > 6$) with more than one Gaussian with very similar mean values. As already stated we do not interpret this as a sign of multiple populations but rather that the tails on both ends of each bin are not captured by a single Gaussian. The exception is the bin around 25-34 days with respect to peak which shows 3 Gaussians for the best fit which do not share the same mean value. Thus we find no evidence for a pre-explosion outburst in days -30 to -15 wrt. maximum light but evidence for populations around the secondary maximum in the R-band.

	We also searched for several populations in the light curve stretch distribution. Again, we used Gaussian mixture models and examined if the fit is improved compared to a single Gaussian fit. 

	\begin{figure*}
		\centering
		\includegraphics[width=14cm]{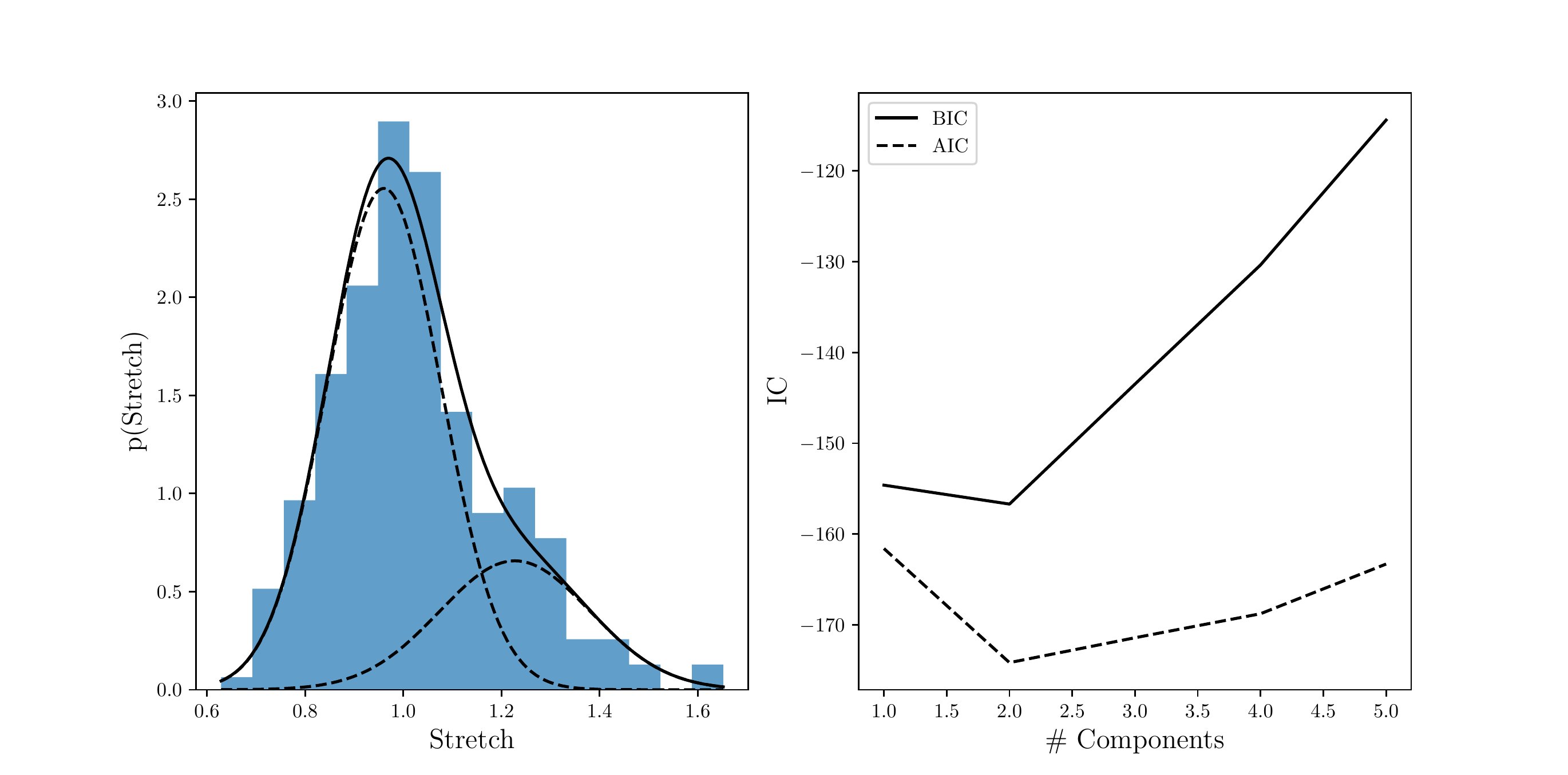}
		\caption{The left panel shows the combined mixture model in a solid line and two individual components in dashed lines. The right panel shows the information criteria (IC): AIC and BIC for different number of Gaussian components. The Gaussian Mixture model fit of the stretch distribution, where we see that both BIC, in the solid line, and AIC, in dashed line, favours two components over one.}
		\label{fig:pop_s}
	\end{figure*} 
    
Figure \ref{fig:pop_s} shows the stretch distribution and the Gaussian mixture model fits, where we find that two Gaussian fit better than one ($\Delta BIC = 2$). We thus conclude that there is no significant evidence for two populations over one. There are many examples in the literature of populations and asymmetry in stretch and colour \citep[e.g.][]{2006AJ....131..527J, 2009ApJ...704..629M, 2011ApJ...731..120M, 2011PhLB..695....1L, 2015AJ....150..172K, 2016MNRAS.460.3529A, 2016ApJ...822L..35S}.

\subsection{Brightness evolution with redshift}\label{Section:zevolution}
	By performing a two-sided Kolmogorov-Smirnov (KS) test on the ``pull distribution", i.e., the error-weighted distribution of estimators around the true value on the binned light curves of different redshifts (seen in Figure \ref{fig:lc_aligned}), we find that the p-values are in many cases lower than 1\%, i.e., we find no significant evidence for evolution in the light curve with redshift at any epoch. If the p-value is zero, it means that we cannot exclude the possibility that the distributions are different. This conclusion holds independent of the choice of bins.

\section{Characterizing the light curve properties}
	In the next Section  we use the unique history of upper limits before the supernova explodes to examine if there are any pre-explosion eruptions or post-explosion flares. Finding a pre-explosion eruption could give information about the progenitor of type Ia SNe. We are able to set limits for such an explosion but do not have the depth to exclude pre-explosion eruptions at the brightness level of a classic nova. We will also examine the average light curve parameters and look for multiple populations within the rise times.
	
	\subsection{Pre- and post- explosion limits}\label{Section:prelims}
	Since our dataset spans many days before and after explosion it is possible to look for pre- and post-explosion eruptions or bumps, similar to novae, which in turn would give us information about the progenitor of SN Ia and possible interaction with the environment of the SN. This was done for type IIn SNe in \citet{2014ApJ...789..104O}. By comparing the history of all individual light curves we looked for bumps before -30 days, and after +200 days with respect to maximum light. We used only the limits that were 20 magnitudes or deeper in this analysis. We do not find any significant perturbations before or after the supernova light is visible.  
	
	\begin{figure*}
		\centering
		\includegraphics[width=\hsize]{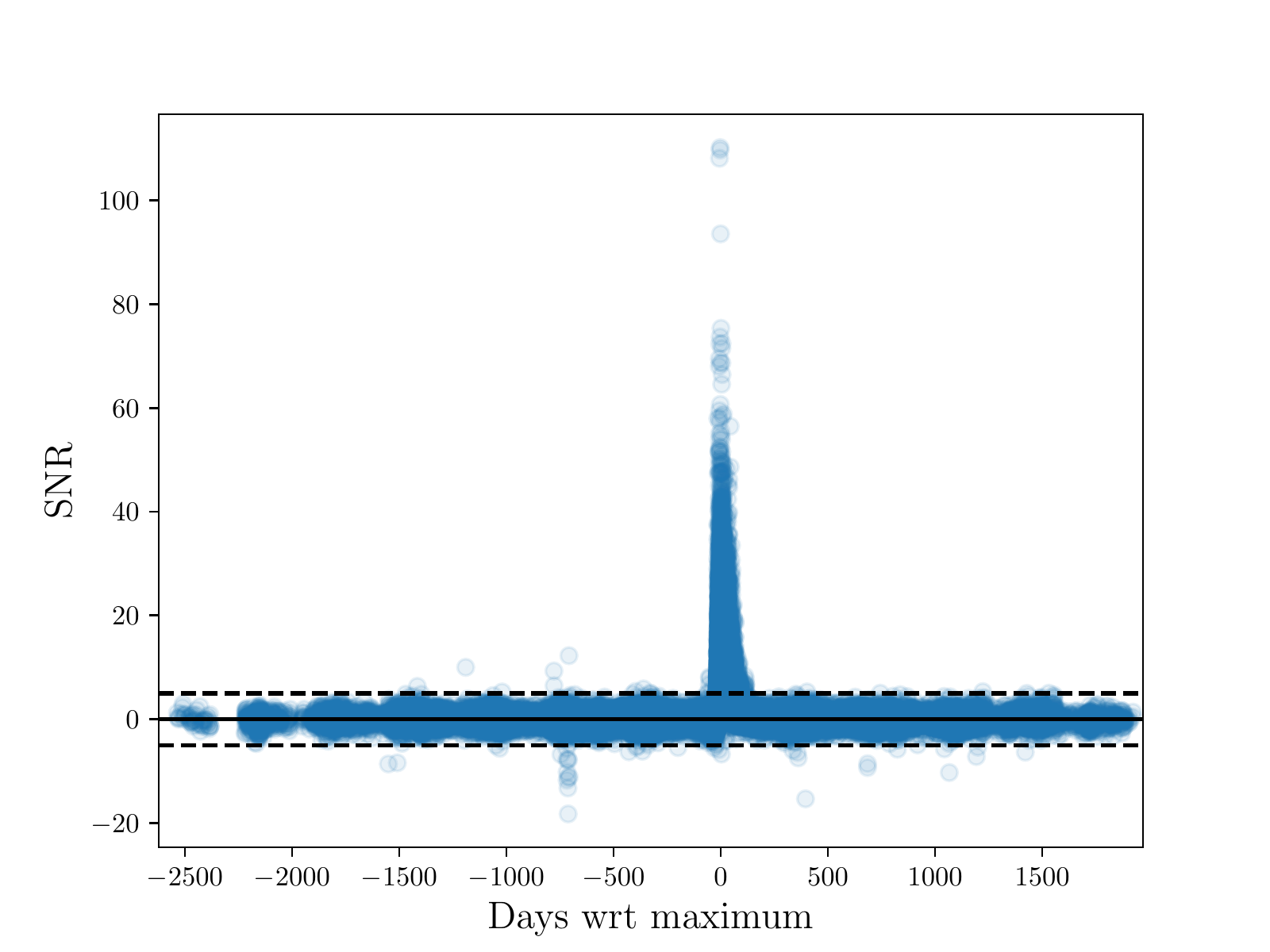}
		\label{fig:limits_abs}
		\caption{Signal-to-noise (SNR) distribution as a function of time from light curve peak of the fluxes of the SNe of our sample. The dashed lines show the 5 $\sigma$ limits. As discussed in the text the deviating data points (that are not part of the light curve, from day -20 to +100) come from various SNe and are not significant.}
		\label{fig:limits}
	\end{figure*}
    
	This might not be surprising since we do not have the sensitivity to detect bumps corresponding to the brightest observed novae, even for the most-nearby SNe in our sample. In Figure \ref{fig:limits}, we show the signal-to-noise ratio of our data points with respect to time of maximum (t=0 in the plot). We are not sensitive to novae since their absolute magnitude range is between $-10$ to $-5$ mag, as shown in \citet{2011BASI...39..375K}.  We report that no eruption brighter than about $-15$ absolute R-band magnitude was found. The deepest limits come from the nearby supernova SN\,2014J \citep[iPTF14jj, see][]{2014ApJ...784L..12G}, showing the strength of nearby supernovae for this type of search. 
	
	Note that the detections in Figure \ref{fig:limits}, outside of the SN region, are not consecutive and thus considered in this analysis as noise. There are a variety of possible explanations for these detections including astrometric errors, cosmic rays, CCD ghosts, variable cloud coverage, other artefacts, unknown asteroids, etc. \citet{2016ApJ...830...27Z} showed that the classical method
for image subtraction underestimates the noise due to several reasons
(e.g., astrometric noise, source noise, correlated noise, reference image noise),
and are less sensitive to cosmic rays \citep[see example in][]{2016ApJ...824....6O}.
    
    We therefore set the criteria to require at least 2 consecutive detections in order to further examine if this is due to a pre-explosion eruption.
	In one case, iPTF13ccm, we observe two consecutive pre-explosion detections at -1000 days with respect to maximum light. This supernova is located near a bright star and thus these detections need to be confirmed. Therefore we run this supernova through an additional photometric pipeline but found that the images were of poor quality and could not confirm a pre-explosion detection. We therefore choose not to trust this pre-explosion detection.

    A deeper survey such as the Large Synoptic Survey Telescope, \citep[LSST][]{2008arXiv0805.2366I}, would be needed to set more stringent limits on pre-explosion eruptions. We note in addition, that we find no post-explosion eruptions in our data. 
	
	\subsection{Early light curves}\label{Section:lcfitting}
\begin{figure}
		\centering
		\includegraphics[width=9cm]{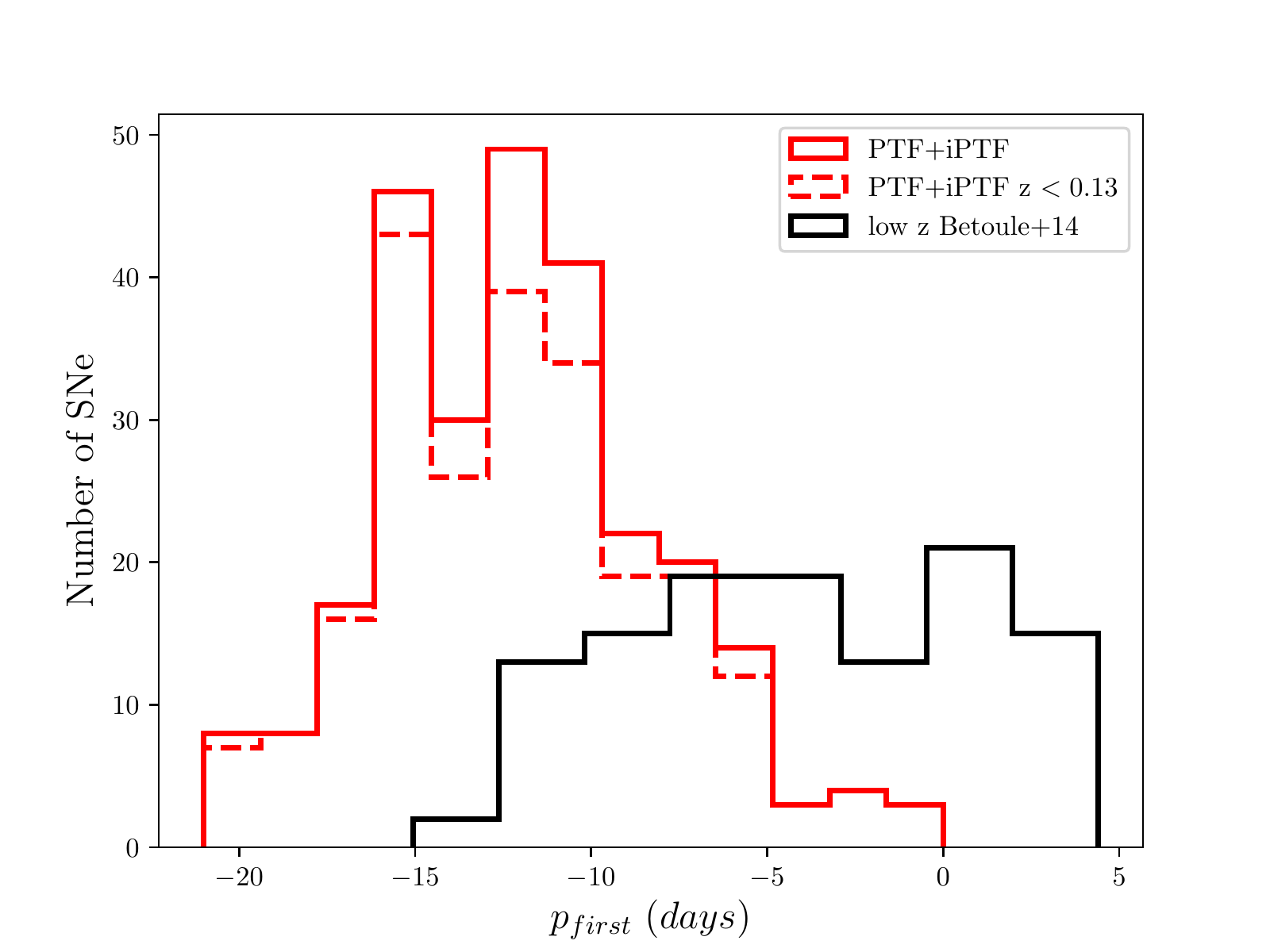}
		\caption{Histogram of earliest detection point, $p_{first}$ in days of our data sample from PTF/iPTF, compared with the low redshift sample from \protect \cite{2014A&A...568A..22B}.}
		\label{fig:first_detection}
	\end{figure}

    The PTF and iPTF sample is unique in that it discovers supernovae very early, compared to other surveys. Comparing the first detection point, $p_{\rm first}$ in our sample with the low redshift literature supernovae from the JLA sample \citep{2014A&A...568A..22B}, we find that the mean changes from $-12\pm3$ to $-4\pm5$ days. This is also illustrated in Figure \ref{fig:first_detection}. The PTF and iPTF sample have data points much earlier on average than the low redshift JLA sample and is therefore well suited for studies of the early part of the light curves. 
    
    Since the 1980's there have been many studies of the early light curves of type Ia SNe. These studies found a correlation between the rise-time of a supernova and its brightness at maximum light, a shorter rise-time corresponding to a less luminous peak brightness. 
	
	While the early studies, \citep[e.g][]{1984SvA....28..658P, 1993ApJ...413L.105P, 1997ApJ...483..565P} were only able to investigate this correlation, later studies with larger and more frequently sampled datasets \citep[e.g.][]{2006AJ....132.1707C, 2007ApJ...671.1084S, 2010ApJ...712..350H, 2011MNRAS.416.2607G, 2012ApJ...745...44G, 2015MNRAS.446.3895F} looked in addition at the parametrisation and shape of the rise. 

\citet{2010ApJ...708.1025K} showed that if SNe Ia originate from a single degenerate scenario, i.e. with a giant companion, in about 10\% of the cases there would be observational evidence of this in the early light curve in the form of an excess of flux. \citet{2010ApJ...712..350H} and \citet{2011MNRAS.416.2607G} found, in their studies of 108 and 61 supernovae light curves respectively, no evidence of interaction with a companion star. While they looked at the stacked light curves we will here examine each light curve individually and parametrise its rise-time and explosion time and then examine the average properties. 
   
	We used the analytical equation presented in \citet{2017ApJ...838L...4Z} to fit our supernovae light curve data to more easily be able to compare our results with literature values instead of using the Gaussian-processes template only. This analytic expression is derived from the photospheric-velocity-evolution function and makes the assumption that the emission is photospheric. It differs from the previous fitting methods by being less sensitive to where there is data in the light curve,  \citep[e.g., compared to][which we found to not be robust for the majority of the light curves in our data set]{2015MNRAS.446.3895F}. We show the results of fitting the analytical equation to our data in Section \ref{Section:lcfitting}.
    
	Now, looking at the individual light curves instead of the sample as a whole we chose to use the empirical equation from \citet{2017ApJ...838L...4Z}, shown in equation \ref{eq:zheng} to fit our light curves in order to obtain parameters, primarily from the early time of the light curve. As mentioned earlier this part of the light curve is potentially important to probe the explosion mechanism and to distinguish between different progenitor scenarios. As opposed to most other empirical fits this equation fits the entire light curve and uses all available data, removing the need to cut at an arbitrary flux level before maximum light such as that used by \citet{2015MNRAS.446.3895F}. The light-curve fits based on \citet{2017ApJ...838L...4Z}, SALT2 \citep{2007A&A...466...11G} and the GP template yield very comparable results, as discussed in Appendix \ref{Section:Comparison}.
	
	The parameters in the equation are the normalising factor $A'$, the explosion time $t_0$, the break time $t_b$, two free parameters determining the shape of the light curve, $\alpha_r$, $\alpha_d$ and a smoothing parameter $s$. 
	
	\begin{equation}\label{eq:zheng}
		L = A' \left[ \frac{t-t_0}{t_b} \right]^{\alpha_r} \left[ 1+\left(\frac{t-t_0}{t_b}\right)^{s \alpha_d}\right]^{\frac{-2}{s}}
	\end{equation}

\begin{figure*}
		\centering
		\includegraphics[width=\hsize]{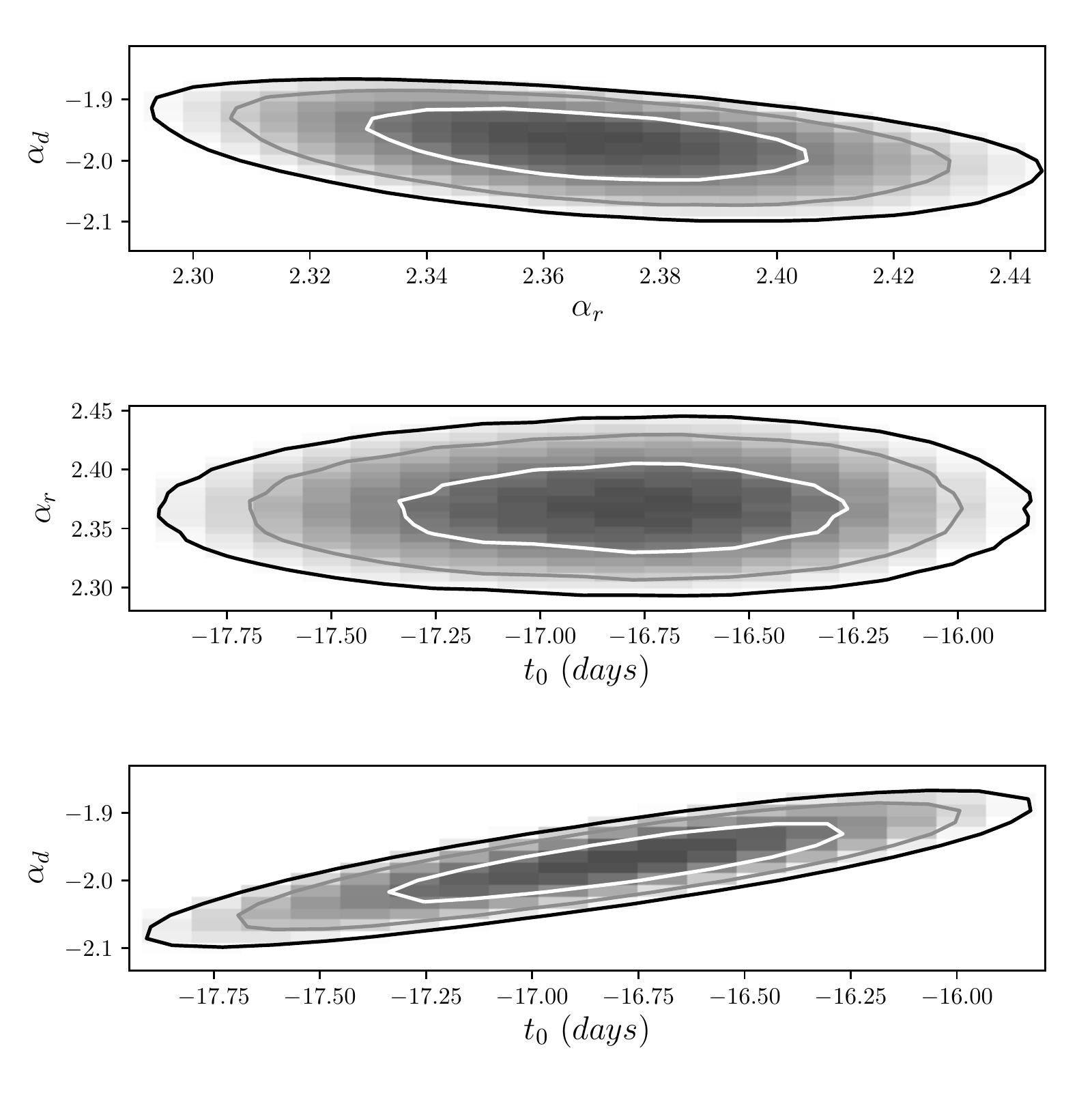}
		\caption{These three panels show the best fit values of equation \ref{eq:zheng} to 207 of the SNe in our sample. Because of the degeneracy between the parameters $t_0$ (in days), $\alpha_d$ and $\alpha_r$ we keep in one of these parameters fixed while the other two are free. 
        The contour lines show 1,2 and 3 $\sigma$ confidence intervals for the sample.}
		\label{fig:contour}
	\end{figure*}

As suggested by \citet{2017ApJ...838L...4Z}, we fix the values of $t_b =20.4$ days. We note considerable degeneracies between several other of the fitted parameters, especially between $t_0$, $\alpha_d$ and $\alpha_r$. The degeneracy is stronger in the cases where data around the rise time is sparse. We show in Figure \ref{fig:contour} the combined limits for all SNe fitted, in total 207, since not all the SNe in the sample have sufficient data points before maximum light to get a good fit, keeping one of the parameters ($t_0$, $\alpha_d$ and $\alpha_r$) fixed at a time. We find the best fit values to be $-16.8^{+0.5}_{-0.6}$ days, $1.97_{-0.07}^{+0.05}$  and $2.36_{-0.03}^{+0.05}$ for $t_0$, $\alpha_d$ and $\alpha_r$ respectively, where the errors stated are the 1 $\sigma$ contours for each respective parameter. The value of the equivalent of $\alpha_r$ can be compared to the other studies which find a value between $\approx 1-3$ \citep[e.g.][]{2006AJ....132.1707C, 2011MNRAS.416.2607G, 2015MNRAS.446.3895F, 2017ApJ...838L...4Z, 2017ApJ...848...66Z} and while it is comparable with other surveys it is higher than expected from a fireball model where $\alpha_r = 2$. We encourage testing different models for this early light curve data.

\subsection{Multiple populations in the rise-time}\label{Section:populations_rise}
As with the stretch distribution we examined the possibility of multiple populations in the fitting parameters of equation \ref{eq:zheng}. We perform Gaussian Mixture models (GMM) on a bootstrapped sample of our data where $\alpha_d$ is kept fixed and search for evidence of multiple populations in the $t_0 - \alpha_r$ parameter space and find no statistically significant evidence for several populations.  We note that the location of the minimum of each individual SN ellipse is widespread but with large errors. Due to these large errors Gaussian Mixture models cannot be used to distinguish possible multiple populations in the data. 49\% of our 1000 bootstrapped samples showed one component fit the data significantly better (with $BIC >6$), 29\% showed 2 components were a better fit and the rest were best fitted with more than 2 Gaussian components. We used the Bayesian information criterion since it sets more stringent restrictions and thus is more suitable to determine if there are more than one population in the data. 

\begin{figure*}
		\centering
		\includegraphics[width=\textwidth]{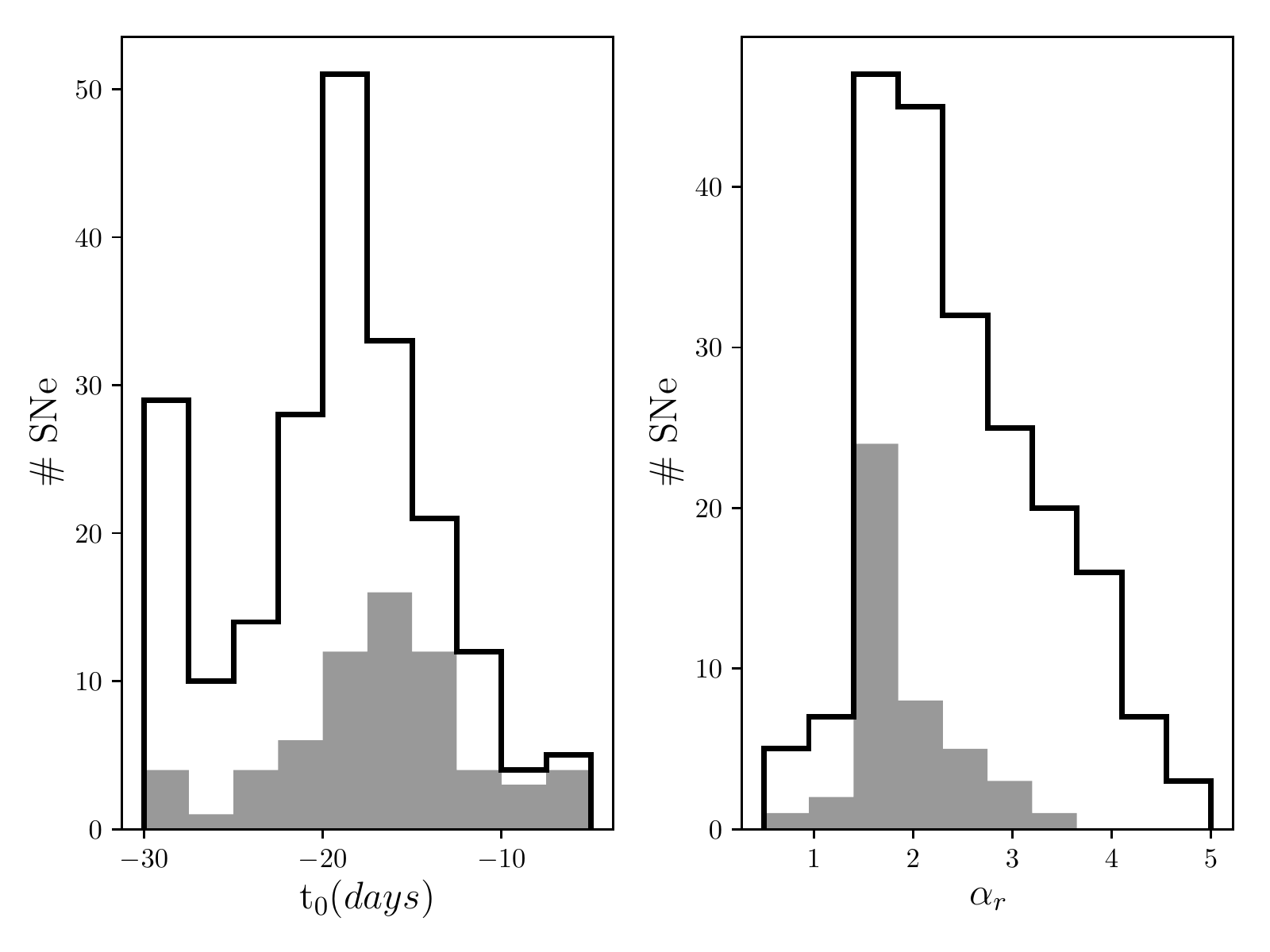}
        \caption{The histograms of the distributions of the best fit values of the $t_0$ and $\alpha_r$ parameters vs number of SNe. The peak at $t_0 \approx -30$ days is driven by SNe with insufficient data points in the early part of the light curve and the error ellipses on these values are sometimes very large, for more details see in the text. The shaded regions show the fits with errors in $t_0 < 2$ days and $\alpha_r <0.2$.}
		\label{fig:pop_rise}
	\end{figure*}

See Figure \ref{fig:pop_rise} for the histograms of the parameters. Note the spike at $t_0 \approx -30$ days in the right panel of Figure \ref{fig:pop_rise} which is driven by SNe with insufficient data points in the early part of the light curve. As seen in the table \ref{table:fitval} in appendix \ref{table:fitvals} many of the best fit parameters have large errors. The fits to the light-curves and their $\chi^2$ can be found in the Supplementary materials. We do not interpret this spike as a hint of a second population, but rather problems with the fitting degeneracy. If more than one population was found this would have pointed towards more than one sub-population of SNe with different progenitor origins.

\section{Examining the Hubble-Lema\^{i}tre residuals} \label{Section:hubble}
\begin{figure*}
		\centering
		\includegraphics[width=\hsize]{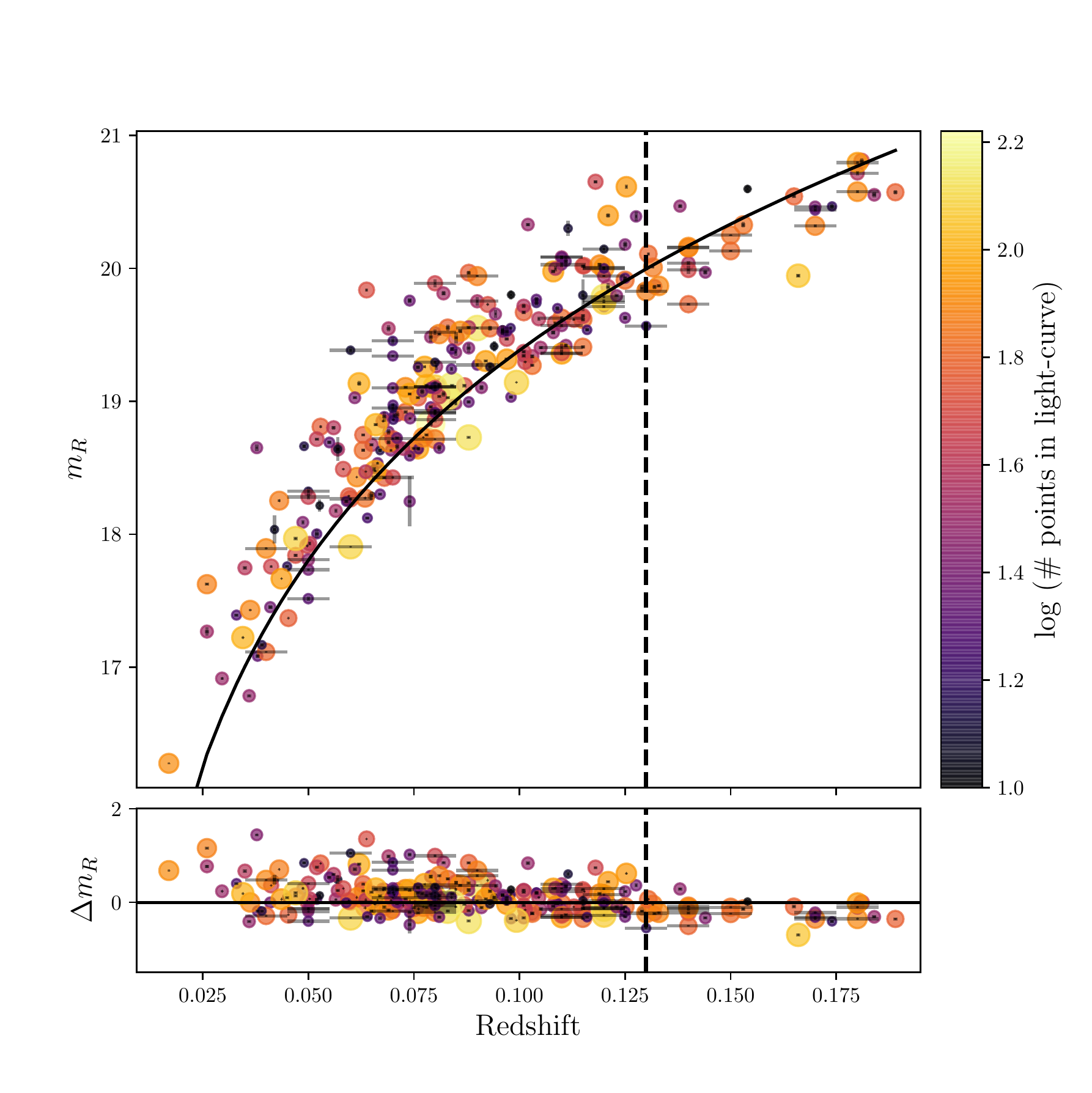}
		\caption{In the top panel we show the Hubble-Lema\^{i}tre diagram, where the size of the data points are scaled logarithmically according to the number of data points that their light curves contain. The solid line shows the standard $\Lambda$CDM cosmology. The Hubble-Lema\^{i}tre residuals for the sample are shown in the lower panel, with the dashed line indicating the redshift at which Malmquist bias becomes important. We do not include the outlier supernova SN2014J, since this supernova is highly reddened and very nearby. As discussed in the text, these SNe are not corrected for extinction.}
		\label{fig:hubbleresiduals}
	\end{figure*}

Using the template as described in Section \ref{Section:template} we get the time of maximum estimate in the R-band for our sample with an accuracy of $\sim$ 1 day. The peak magnitude is then plotted against redshift in a Hubble-Lema\^{i}tre diagram and shown in Figure \ref{fig:hubbleresiduals}. The rms of the Hubble-Lema\^{i}tre residuals is 0.35 magnitudes for all redshifts after stretch corrections. In section \ref{Section:extinct} we discuss our estimate of the uncertainty stemming from not being able to correct for extinction. Figure \ref{fig:dist2} shows that this can be quite large, with a tail reaching $>0.5$ mag.
	
	\subsection{Malmquist bias}\label{Section:malm}
    An important systematic for type Ia SN cosmology is Malmquist bias \citep{1922MeLuF.100....1M}, which is the redshift on beyond which the survey becomes flux limited, i.e. when we probe only the brightest SNe rather than the entire population. We determine at which redshift this bias becomes important for our sample in order to account for this and to plan future survey strategies for the Zwicky Transient Facility (ZTF). We thus need to estimate the underlying distribution of Hubble-Lema\^{i}tre residuals. To do this, we fit the convolution of two functions, an exponential and a Gaussian to estimate the mode at different redshift bins. 
	
\begin{figure}
		\centering
		\includegraphics[width=\hsize]{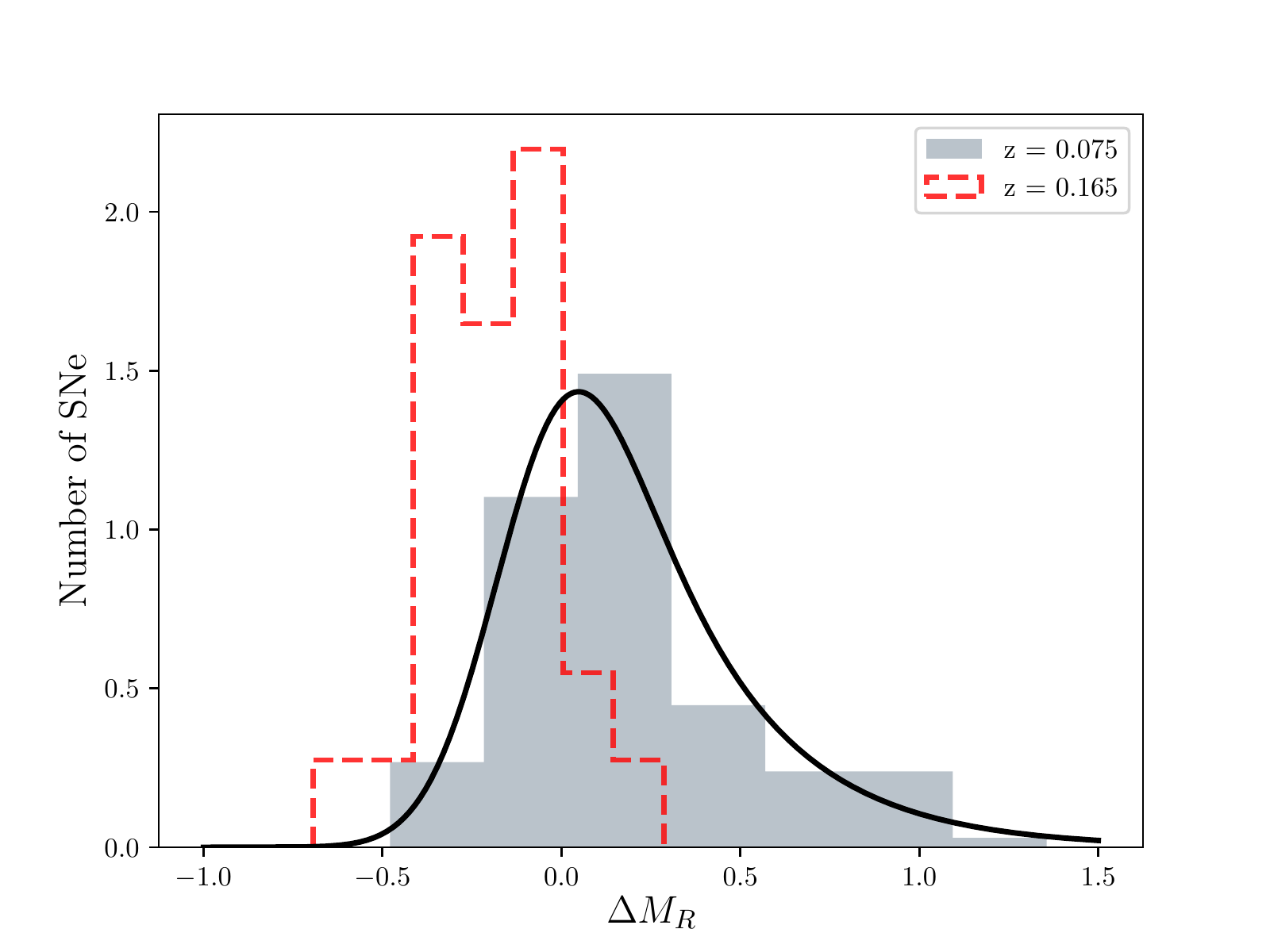}
		\caption{We show the distribution of Hubble-Lema\^{i}tre residuals for two different redshift bins centred around 0.075 and 0.165 in shaded grey and the dashed line respectively. The solid black line shows the best fit convolution between a Gaussian and an exponential used to determine the mode of the histograms of Hubble-Lema\^{i}tre residuals in order to estimate where the Malmquist bias becomes important.}
		\label{fig:exampleyellow}
	\end{figure}

	To determine where the Malmquist bias becomes important we require a 3 $\sigma$ deviation in the Hubble-Lema\^{i}tre residuals. This is found at both high and low redshifts. At low redshifts the mode is 3 $\sigma$ above zero due to peculiar velocities and highly extinct SNe at low redshift.  At higher redshift, we can see that we get a 3.4 $\sigma$ deviation to the faint end at $z = 0.13$. In Figure \ref{fig:hubbleresiduals} the dashed line shows where this limit lies in the Hubble-Lema\^{i}tre diagram and in Figure \ref{fig:exampleyellow} we show the histogram of two bins, one of which is Malmquist biased. We thus determine that Malmquist bias becomes statistically significant at redshift 0.13 for our sample. 
    
	\subsection{Average extinction and mean dust path}\label{Section:extinct}
    One of the largest systematic of type Ia SNe is the extinction by dust. This can be corrected for using the colour-magnitude correlation found in literature.

	Since our sample does not have additional filter information, this correction could not be performed for individual SNe, however we were able to estimate the average path length of dust that the SN light travelled through for our sample. This can then be translated into an average extinction of all SNe in our sample to correct the maximum magnitude of R-band SNe.
    
    To understand the origin of the Hubble-Lema\^{i}tre residual distribution we use the SuperNova Observation Calculator \citep[SNOC, described in][]{2002A&A...392..757G}, to create simulated supernova samples with different amounts of extinction. We use the code to generate samples of $2000$ type Ia SNe using the same redshift distribution we have from our iPTF and PTF sample.
	
	For each iteration we change two parameters; the intrinsic scatter (characterised by the width of the Gaussian part in fitting the Gaussian convoluted with an exponential as we did to determine the Malmquist bias) and the mean free path for host galaxy dust extinction. We allow the values to vary from $0.1-0.30$ magnitudes and $1 \times 10^{-5}-1 \times 10^{-2}$ Mpc for intrinsic scatter and host dust extinction respectively. We then compare the Hubble-Lema\^{i}tre residual distribution from each $SNOC$ iteration with our own sample distribution using a double-sided K-S test.
	
    We find the minimum to lie at 1 kpc corresponding to a mean $E(B-V)$ of $\approx 0.05(2)$ magnitude \footnote{The number in parenthesis denotes one standard deviation from the mean.} or an $A_R \approx 0.11$ magnitude, assuming $R_V = 3.1$.
 	
	\begin{figure}
		\centering
		\includegraphics[width=\hsize]{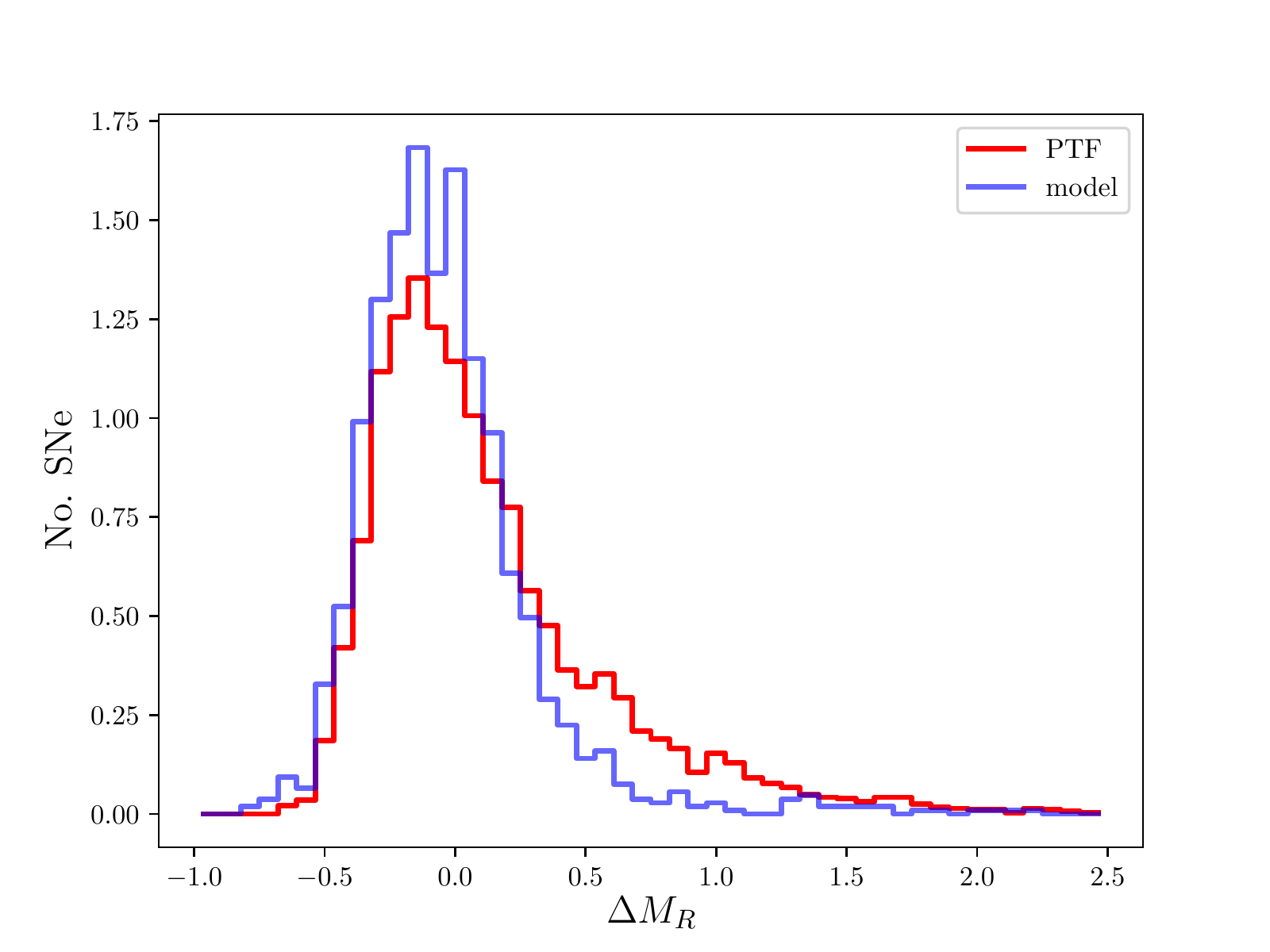}
		\caption{An example plot of the normalised Hubble-Lema\^{i}tre residuals, $\Delta M_R$ where the SNOC model is consistent with our distribution. The red curve shows the PTF and iPTF Hubble-Lema\^{i}tre residual distribution and the blue curve the model with a mean free path of $1$ kpc and an intrinsic scatter of $0.2$ magnitudes. The plot is normalised so that the area under each curve equals 1.}
		\label{fig:dist2}
	\end{figure}
	
	While the double sided K-S test does not give a confidence interval the results are consistent with an average mean free path of $10^{-3}$ Mpc. An example where the model is consistent with the Hubble-Lema\^{i}tre residuals in our sample is shown in Figure \ref{fig:dist2}. It is important to note that the SNOC simulations are idealised and treat measurement errors in a simplified way, thus we do not get a very good fit to our data. 
    We do not reach a clear minimum for the intrinsic scatter parameter. By visual examination of the fits the negative Hubble-Lema\^{i}tre residuals are overestimated for high values of intrinsic scatter in the model, yet yield a lower K-S statistic. While this means that we cannot constrain the intrinsic scatter using this method, the common minimum at $1$ kpc for all values of the intrinsic scatter suggests that the average mean free path we get is consistent with our data. The intrinsic scatter is thus constrained using the Gaussian part of the fit to the convolution of a Gaussian and an exponential (which was also used to obtain the Malmquist bias) and is found to be $0.186 \pm 0.033$ magnitudes for the redshift range 0.05 to 0.1.
    
    From these results we have a better understanding of the average bias that our Hubble-Lema\^{i}tre residuals have since they have not been corrected for colour. 
	
	We attempted to use the low-resolution spectra taken to classify the SNe (at least one per supernova) to get an estimate of the amount of extinction. However synthetic colours do not show any correlation with Hubble-Lema\^{i}tre residuals and thus cannot be used to correct for extinction. This is thought to be due to the uneven flux calibration performed on these spectra. This was also noted by \cite{Maguire14} for the PTF spectra. Note that we do not correct for gravitational lensing of objects in the line of sight in the simulations. This effect is negligible at the these low redshifts. 

\subsection{Mass step in SN hosts}\label{Section:mass}
The aim is to examine the correlation between the host mass and Hubble-Lema\^{i}tre residuals found in several papers with varying degrees of significance on the slope in the B-band \citep[e.g.][]{2010ApJ...722..566L, 2010MNRAS.406..782S, 2013ApJ...770..108C, 2016ApJ...821..115W, 2014MNRAS.438.1391P,2010ApJ...715..743K, 2017arXiv171000845S, 2018arXiv180505911J,2018arXiv180603849R}.

\begin{figure}
		\centering
		\includegraphics[width=\hsize]{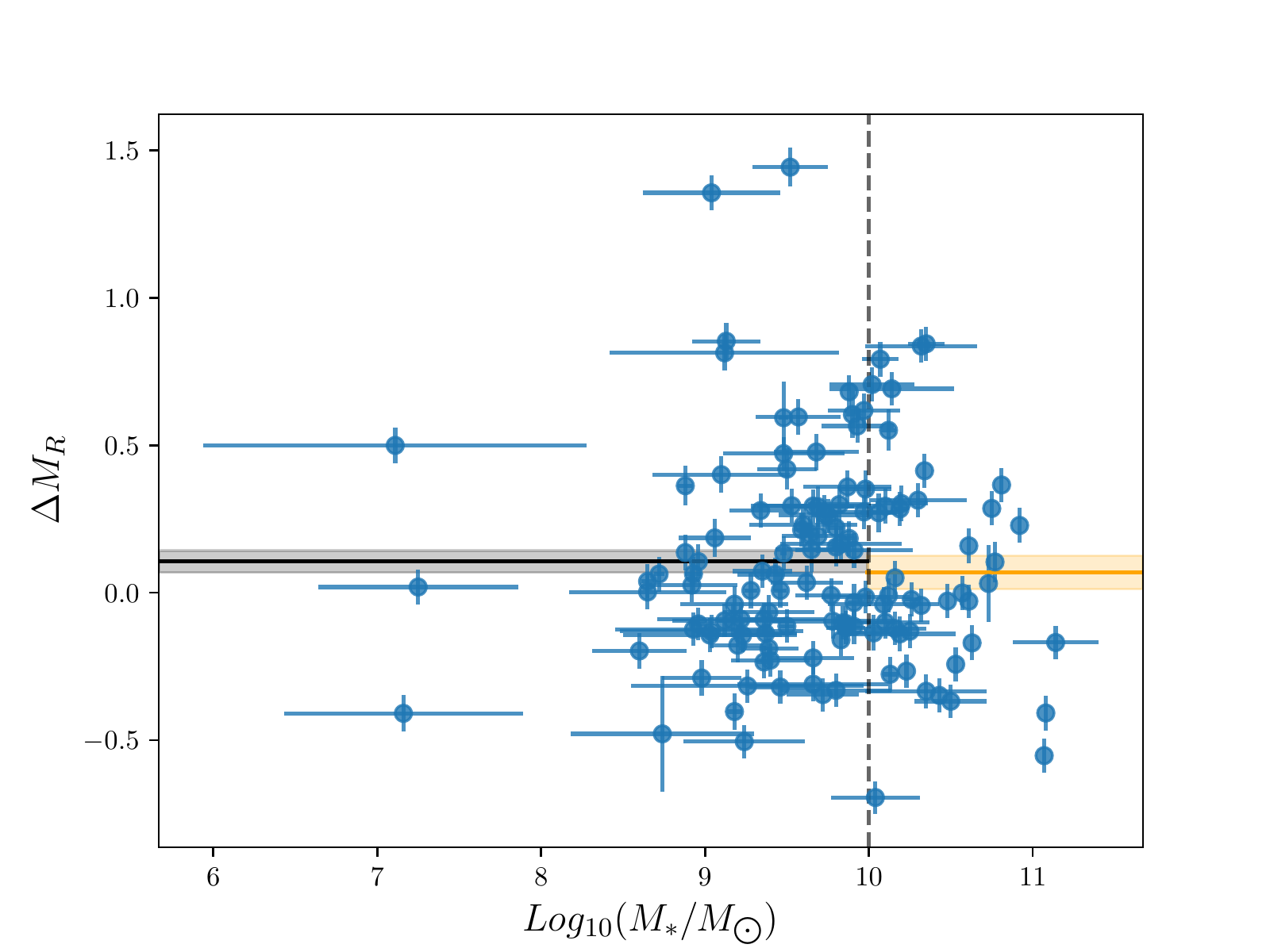}
		\caption{We are showing the Hubble-Lema\^{i}tre residuals in R-band, $\Delta M_R$ vs. the log of the host stellar mass, $\rm log_{10}M_{\ast}/M_{\bigodot}$, for 131 of the SNe in our sample that have reliable host masses. We include K-correction, calibration, photometric and peculiar velocity errors in the Hubble-Lema\^{i}tre residual error-bars. The dashed line shows the definition of high and low mass host galaxy  \protect\citep[see e.g.][]{2010MNRAS.406..782S, 2011ApJ...737..102S}, and the horizontal lines with the shaded areas show the mean and standard error for each of the two host mass bins.}
		\label{fig:mstep}
	\end{figure}

We show in Figure \ref{fig:mstep} the Hubble-Lema\^{i}tre residuals in the R-band from our sample with $z<0.13$ and the log mass of the host galaxies from Hangard et al. (in prep.).

Hosts stellar masses are calculated using FAST \citep[Fitting and Assessment of Synthetic Templates][]{2009ApJ...700..221K}, a code that fits stellar population templates to photometry. We use $ugriz$ magnitudes from SDSS \citep{2015ApJS..219...12A} and $JHK_s$ magnitudes from 2MASS \citep{2006AJ....131.1163S}.
Each host must have a known redshift, and at least 2 data points in magnitudes. Only photometry with errors smaller than 0.25 magnitudes are considered. The stellar populations library used is FSPS by \citet{2010ascl.soft10043C}, and the star formation history is chosen delayed, exponentially declining. The initial mass function is from \citet{2003PASP..115..763C}, and the dust law is from \citet{2013ApJ...775L..16K}. The metallicity is fixed to solar metallicity value (Z = 0.019). We only keep the fits for which the reduced $ \chi^2$ is smaller than 2.

We find the Hubble-Lema\^{i}tre residual step is $0.037 \pm 0.068$ is compatible with the latest results from \citet{2018ApJ...859..101S}. However, our results is also compatible with no step in the Hubble-Lema\^{i}tre residuals.  We found no redshift dependence on the mass step measurement for $z \le 0.13$, which is why we restricted the SNe to that redshift range, coinciding with our adopted estimate of the onset of a significant Malmquist bias, see section \ref{Section:malm}.

	\section{Discussion} \label{Section:discussion}

 We presented the light curve analysis from PTF and iPTF, an un-targeted survey which addresses one of the main problems in present day cosmology with type Ia SNe; namely the sampling bias. However, since we do not address another significant bias, the colour of the SNe, we have focused this paper on looking at the average light curve properties. 
   
    \begin{figure*}
		\centering
		\includegraphics[width=15cm]{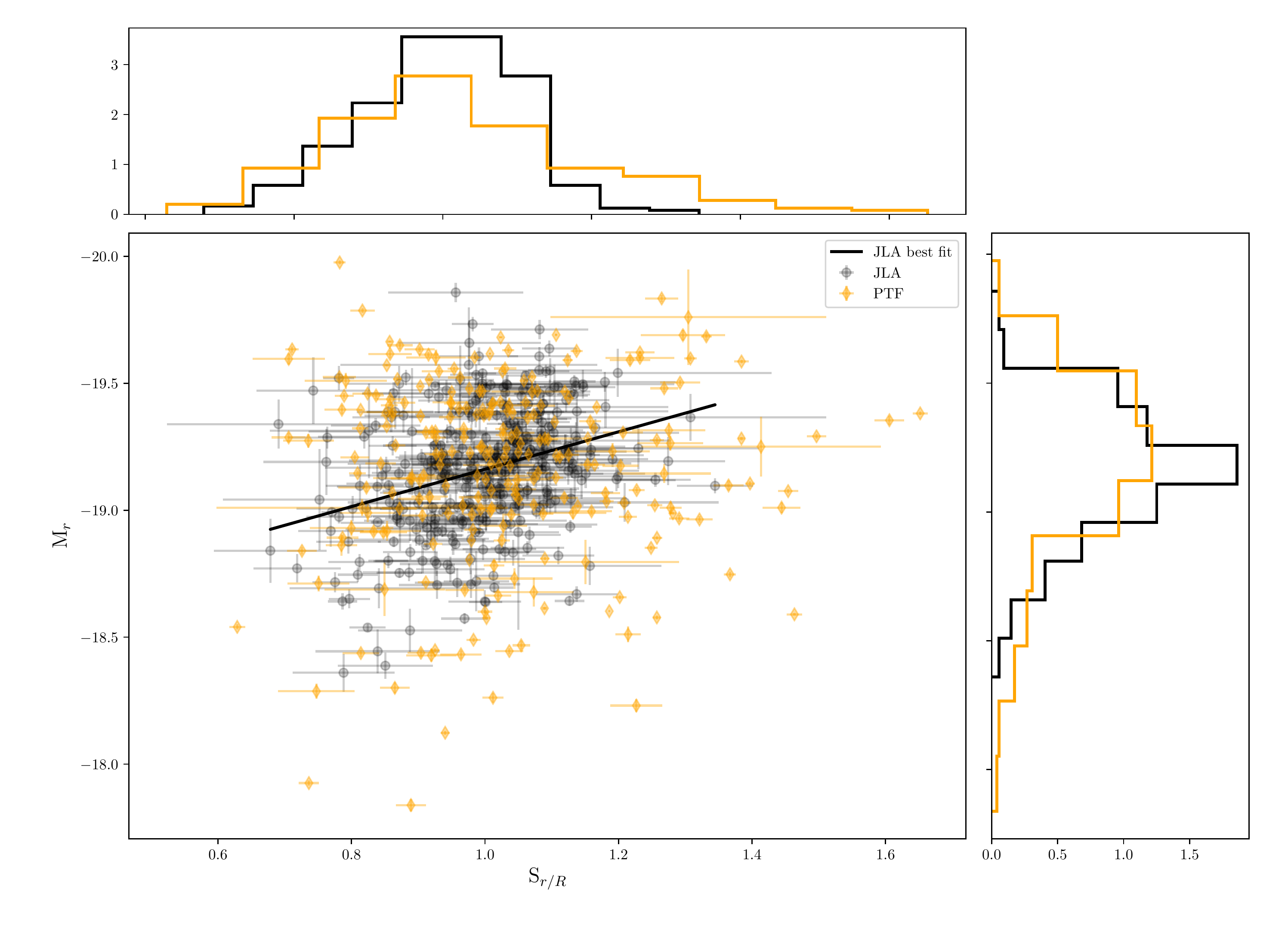}
	\caption{Showing the peak absolute magnitude, $M_r$ vs. stretch, $S_r/R$ relation for the JLA nearby supernova sample, \protect\citet{2014A&A...568A..22B}, in the SDDS r-band in black circles and the PTF and iPTF sample in orange. We also show the best fit line the JLA sample, showing the weak but significant correlation between the parameters. For the PTF sample this correlation is weaker. Note that we have performed an offset corresponding to the S-correction, \protect\citet{2005PASP..117..810S}, of 0.35 magnitudes between the two different filter bands.}
		\label{fig:deltam15}
	\end{figure*}
    
   A commonly used way to reduce the Hubble-Lema\^{i}tre residuals is to use the relation between the peak brightness and the width of the light curve, such as stretch \citep{1997ApJ...483..565P}. In order to compare with literature on r-band only fits we used \texttt{sncosmo} to calculate the absolute magnitudes and stretch of the JLA low redshift supernova sample from \citet{2014A&A...568A..22B} using the template from \citet{Hsiao07}. The results for the fits based exclusively on the SDSS r-band are shown in Figure \eqref{fig:deltam15}. To estimate the significance of the correlation between the two parameters $S_R$ and $M_r$ we use Spearman R statistic and bootstrap the data-points according to their individual errorbars and covariance between the two parameters. We do this 10 000 times and find that the average Spearman $R = 0.2$ with $p-value < 10^{-7}$. For the PTF sample the correlation is weaker. If we now compare the slope of this with that of the B-band from \citet{2011AJ....141...19B} (with $\Delta m_{15 B}$) with a value of $0.58\pm0.10$ we see that the slope is less steep in the redder band. This could be due to the relative flatness of the R-band light curve compared to other photometric bands. 
  We also note that, after having performed an S-correction of 0.35 magnitudes, the calibration the PTF and iPTF SNe are consistent with that of the low redshift JLA sample. 
	
	While we in this work look at the average properties of type Ia SNe from an untargeted survey we do not take other biases into account. To improve the quality of this data sample there are a number of things that can be done. Perhaps the most important is to have colour information for each SN such that extinction can be corrected for on an individual SNe level. Secondly, a better calibration of the photometry would be very beneficial. Both these changes are being applied to the ZTF, \citep{2014htu..conf...27B} type Ia SNe survey as well as expanding the data sample. ZTF came online in February 2018 \citep{2018ATel11266....1K} and will be 15 times more efficient than iPTF. With a substantially larger field of view of 47 deg$^2$, faster reading\footnote{Time it takes to read out the data from the camera.} and slewing\footnote{Time it takes the telescope to move from one target to another.} speed it is expected to be able to find 15 times the amount of transient events, including many SNe Ia. Other future surveys of importance for SN Ia discovery and follow-up include the LSST \citep{2008arXiv0805.2366I} which is scheduled to be operational in 2022.	
	
	\section{Conclusion}
    We present in this paper the best 265 sampled SNe type Ia from homogeneous PTF and iPTF dataset in order to examine the light curve properties in the Mould R band of a non-targeted survey. The full tables are in Appendix \ref{table:fitvals} with both the values from the R-band light curve and the individual parameters from the fit of equation \ref{eq:zheng} from \cite{2017ApJ...838L...4Z}. All individual light curve photometry used in this paper is made publicly available through WISeREP\footnote{\url{https://wiserep.weizmann.ac.il}}, \citep{2012PASP..124..668Y}.

    Our conclusions can be summarised as follows:
    \begin{itemize}
    \item We constructed and present a non-parametric template of our sample SNe spanning between -20 and +80 days with respect to maximum light. Since this was constructed with the help of heteroscedastic Gaussian processes we can provide a 90\% confidence region around the template that takes the errors of each data point into account. We used this to examine the intrinsic scatter and found no evidence for multiple populations at any bin along the template. We note a wider spread around the time of the light curve shoulder, $\approx 30$ days after peak.
    \item We determined the Malmquist bias in our sample to become noticeable at $z=0.13$ by fitting a Gaussian and an exponential to the Hubble-Lema\^{i}tre residuals.
    \item Since this survey was made in one band we cannot correct for individual SNe extinction. We thus determine the average extinction to be E(B-V) $\approx 0.05(2)$ magnitudes or $A_R = 0.11$ magnitudes and the average mean free path for dust extinction to be $10^{-3}$ Mpc by comparing to simulations with SNOC.
    \item We find no redshift evolution in the light curve at any epoch in our sample, when dividing into 3 redshift bins, up to z=0.2. 
    \item We search for pre- and post- explosion flares in our data spanning from -2500 days to +2000 days with respect to maximum and find no significant flare. We note that nearby SNe are especially useful in setting these limits and that the PTF/iPTF depth is not enough to reach the brightness of a novae.
    \item We used the analytical equation presented in \cite{2017ApJ...838L...4Z}, equation \ref{eq:zheng} and fit to 200 of our light curves and get a rise time and rise index for each SN. We then look at the average properties of these and found the best fit values to be $-16.8^{+0.5}_{-0.6}$ days, $1.97_{-0.07}^{+0.05}$  and $2.36_{-0.03}^{+0.05}$ for $t_0$, $\alpha_d$ and $\alpha_r$ respectively, where the errors shown are the larger 1 $\sigma$ contours from the contour ellipses of the parameter fits.
    \item We searched for multiple populations using Gaussian mixture models in individual bins around the Gaussian processes template of the light curves, stretch and rise times as measured with equation \ref{eq:zheng}. We did not find significant evidence of more than one population in any of  these parameters.
    \item We find that the Hubble-Lema\^{i}tre residual step is $0.037 \pm 0.068$ which is both compatible with a zero slope and literature values.  We conclude that our data is not sensitive enough to probe the host mass -luminosity relation.
    \end{itemize}
    
    \section*{Acknowledgements}
	SP would like to thank D. Men\'endez Hurtado, K. Muroe, D. Mortlock and T. Calv\'en for helpful discussions. The authors thank the anonymous referee for comments and suggestions which improved the paper.
    
    The Intermediate Palomar Transient Factory project is a scientific collaboration among the California Institute of Technology, Los Alamos National Laboratory, the University of Wisconsin, Milwaukee, the Oskar Klein Center, the Weizmann Institute of Science, the TANGO Program of the University System of Taiwan, and the Kavli Institute for the Physics and Mathematics of the Universe. The iPTF Swedish collaboration is funded through a grant from the Knut and Alice Wallenberg foundation and individual grants from the Swedish National Science Council as well as the Swedish National Space Agency. DAH is supported by NSF grant AST-1313484. 
	This work was supported by the GROWTH project funded by the National Science Foundation under Grant No 1545949. GROWTH is a collaborative project between California Institute of Technology (USA), Pomona College (USA), San Diego State University (USA), Los Alamos National Laboratory (USA), University of Maryland College Park (USA), University of Wisconsin Milwaukee (USA), Tokyo Institute of Technology (Japan), National Central University (Taiwan), Indian Institute of Astrophysics (India), Inter-University Center for Astronomy and Astrophysics (India), Weizmann Institute of Science (Israel), The Oskar Klein Centre at Stockholm University (Sweden), Humboldt University (Germany). The Weizmann interactive supernova data repository - http://wiserep.weizmann.ac.il was used to make the data public. This research was conducted using the resources of High Performance Computing Center North (HPC2N) under the proposal SNIC 2017/3-64. Part of this research was carried out at the Jet Propulsion Laboratory, California Institute of Technology, under a contract with the National Aeronautics and Space Administration.

	
	
	
	\bibliographystyle{mnras}
	\bibliography{references} 
	
	
	\appendix
 \section{Photometric filter}\label{Section:filter}
\begin{figure}
		\includegraphics[width=10cm]{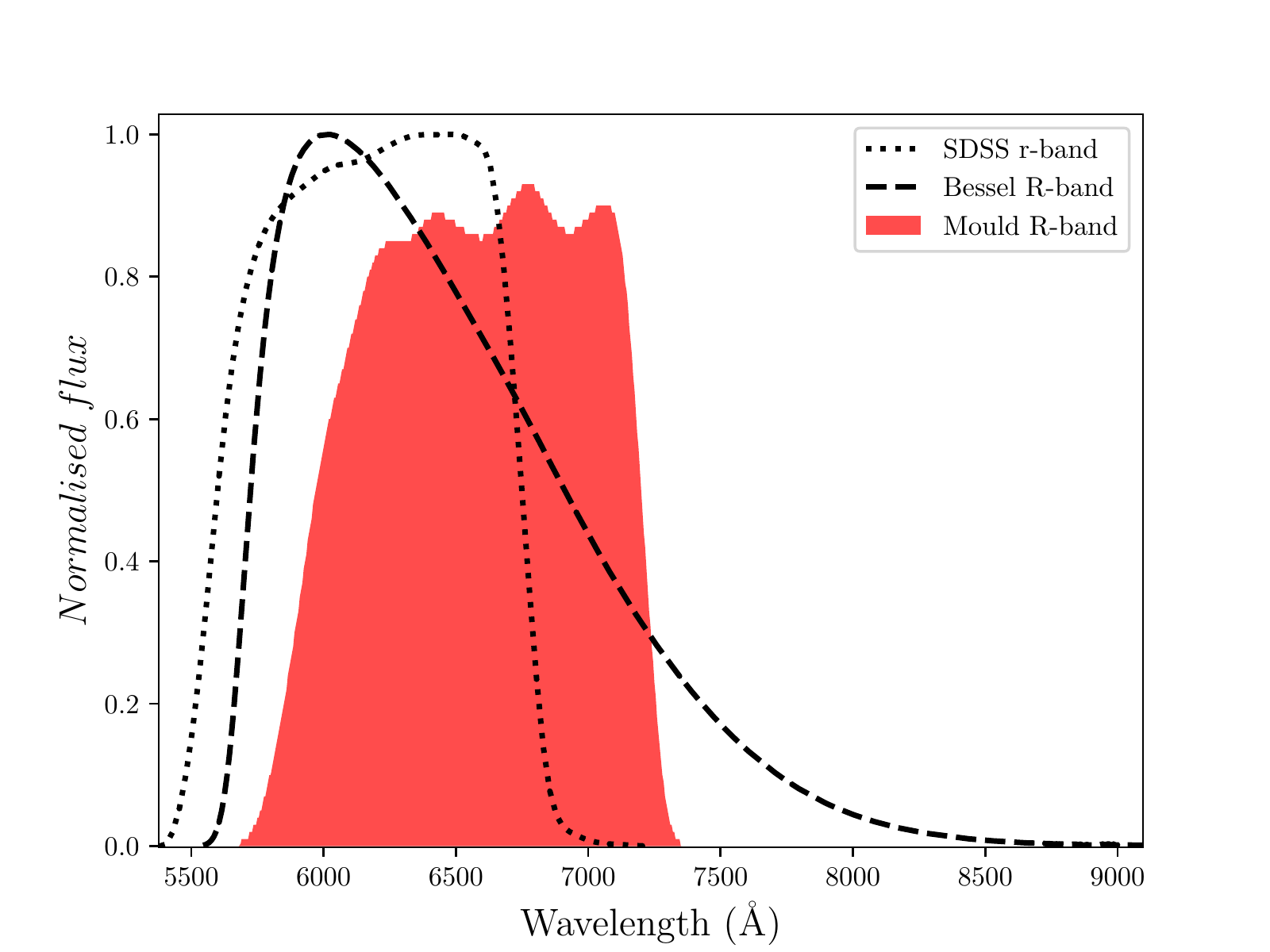}
		\caption{This plot shows a comparison between different filter functions used, we show the Mould R-band used in this paper for our data analysis, the Bessel R and the SDSS r filter.}\label{fig:filter}
	\end{figure}

 In Figure \ref{fig:filter} we show how the R-band filter used for our data sample compares with other filters more commonly used in the literature, such as the Bessel R and SDSS r, see \cite{2005ARA&A..43..293B} for a review of different filters. The latter was used in \cite{2014A&A...568A..22B} to which we compare our sample in Section \ref{Section:discussion}.
    \section{Forced Photometry and magnitudes in our dataset}\label{appendix:forced}
    
\subsection{Baseline correction}
We have used forced photometry in our analysis which is performed with difference imaging of the data and gives a relative photometry. We then convert this to an absolute photometry as described in Section \ref{Section:absolutephot}. Before that conversion we make a baseline correction to the initial light curve to correct for any residual offset in the ``history" of the light curve. We choose to define any point earlier than 50 days before peak to be defined as ``history" and use these points to determine the level of this baseline. The baseline correction is necessary to account for when the reference image was taken. If the reference image includes SN flux or includes a different systematic the photometry will not be correct without this correction. In the light curves accompanying this paper there is a flag for when this baseline correction could not be performed due to lack of sufficient ``historical" data points.

\subsection{Quality checks}
We perform two checks to ensure that the photometry can be trusted. 
\begin{itemize}
\item We compare the point spread function (PSF) photometry to aperture photometry to see if there is any global bias for positive flux measurements and to detect global systematics in the PSF-templates since aperture photometry is less affected by astrometric error.
\item We only use photometry with PSF sharpness (a parameter given by the pipeline) of $\approx 1$ indicating a PSF-like source rather than a spike or extended profile.
\end{itemize}

\subsection{Uncertainties in the photometry}

We calculate the uncertainties in the fluxes by multiplying the 1 $\sigma$ uncertainties in the PSF-fit fluxes with a scaling factor as shown in equation \ref{eq:meth}. 

\begin{equation}\label{eq:meth}
\sigma_{F(\rm corrected)} = scaling \hspace{1mm} factor \times \sigma{F(\rm raw)}
\end{equation} 

The scaling factor is defined as the division of the standard deviation and the median of the ``historic" flux, $F (t_i, t_f)$ as shown in equation \ref{eq:meth2}.

\begin{equation}\label{eq:meth2}
scaling \hspace{1mm} factor =  \sigma_{F (t_i, t_f)}\hspace{1mm} / <F(t_i, t_f)>
\end{equation}
This way of calculating the uncertainties assumes that there is no transient light in the ``historical" part of the light curve. 

\subsection{Absolute photometry}\label{Section:absolutephot}
We then convert the relative photometry to absolute photometry by using the zero point extracted from the reference-image SExtractor catalogue \citep{1996A&AS..117..393B} for stars between the R-band magnitude, $14.5\leq m_R \leq 19.0$ using aperture photometry. If the zero point was not possible to get for a particular image we used the median zero point from the rest of the measurements for the same object. All measurements with a signal-to-noise of more than 3 are classified as detections and thus their magnitude is found with equation \ref{eq:absmags},
\begin{equation}\label{eq:absmags}
M = ZP - 2.5 \log(F(\rm corrected))
\end{equation}
   otherwise we report them as limits following equation \ref{eq:limit}.
   \begin{equation}\label{eq:limit}
M_{limit} = ZP - 2.5 \log(3* \sigma_{F(\rm corrected)})
\end{equation}
	\section{Gaussian processes in machine learning applied to SN light curves} \label{appendix:GP}
	Gaussian processes is a machine learning algorithm for non-parametric regression, i.e. it allows reconstruction of a function without assuming parametrisation or functional form. For a more complete overview of Gaussian processes, see \citet{Rasmussen:2005:GPM:1162254}. We are looking for the latent function (i.e. the true function) $f(t)$ that maximises the likelihood of producing the observed data under the assumption of independent Gaussian noise. Gaussian Processes approximates the latent function as
	\begin{equation}
		GP(m(t), k(t, t')) \approx f(t),
	\end{equation}
	given the expected mean, $m(t)$, and a \textit{covariance function} or kernel, $k(t,t')$, defined to be: 
	\begin{align}
		m(t) &= \mathbb{E}\left[f(t)\right] \\
		k(t, t') &= \mathbb{E}\left[(f(t) - m(t)) (f(t') - m(t'))\right].
	\end{align}
	where $\mathbb{E}$ denotes the expectation value. 
	
	The kernel is a measure of similarity between two points, which can be defined as a distance between two functions $f$ and $g$ as:
	
	\begin{equation}
		d(f, g) = {f|k|g} = \int_{\mathcal{R}}  f(t) k(t, t') g(t') dt dt'.
	\end{equation}
	
	One of the most commonly used kernels is the squared exponential (also called Radial basis function, $RBF$) defined in equation \ref{eq:rbf}, where $\sigma$ is the noise of the data and $l$ the length scale of the kernel. 
	
	\begin{equation}\label{eq:rbf}
		k(t, t') = \sigma^2 \hspace{1mm} e^{-\left(\frac{(t-t')^2}{2l^2}\right)}
	\end{equation}
	The length scale defines the distance between points at which correlation between them is lost. In other words if points are much further away from each other than the length scale they become irrelevant. This kernel depends on two \textit{hyper-parameters}, $\sigma$ and $l$ that have to be set (see Section \eqref{kerneldiff}).
	
	\subsection{Model Section of kernels}\label{kerneldiff}
	The likelihood of obtaining the vector of $N$ observations $\underline{\mathbf{y}} =[y_1, y_2... y_N]$ at points $\underline{\mathbf{T}} = [t_1, t_2... t_N]$ given a kernel of \textit{hyper-parameters} $\underline{\boldsymbol{\theta}}$ (in our squared exponential example, $\underline{\boldsymbol{\theta}}= [\sigma, l]$)  is given by:  
	\begin{equation}\label{eq:likely}
		\log p(\underline{\mathbf{y}} | \underline{\mathbf{T}}, \underline{\boldsymbol{\theta}}) = - \frac{1}{2} \underline{\mathbf{y}}^T K^{-1} \underline{\mathbf{y}} - \frac{1}{2} \log |K| - \frac{N}{2} \log 2 \pi
	\end{equation}
	where the \textit{covariance matrix}, $K_{ij} = k(t_i, t_j)$ containing the pair-wise distances between data points. The first term of equation \ref{eq:likely} measures the goodness of the fit, the second is a complexity penalty and the third is a normalisation. 
	
	\begin{figure}
		\includegraphics[width=10cm]{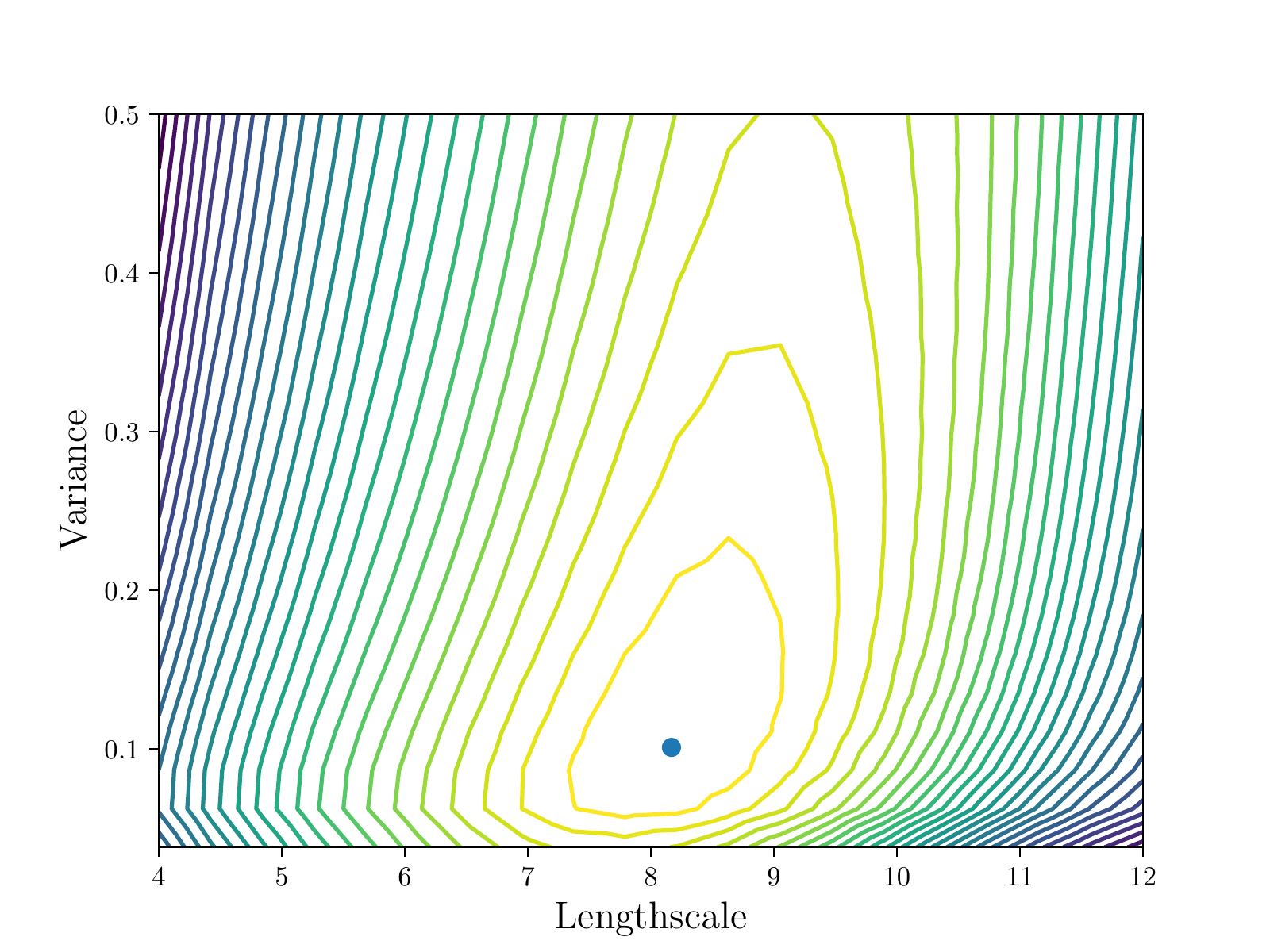}
		\caption{A contour plot of the log likelihood as a function of length-scale and variance \textit{hyper-parameters} of the kernel for a light curve from our sample, iPTF13asv. The dot marks the optimal choice of \textit{hyper-parameters}.}\label{fig:contl}
	\end{figure}
	
	The gradients of equation \eqref{eq:likely} with respect to the \textit{hyper-parameters} can be computed analytically; so we can efficiently compute the \textit{hyper-parameters} that maximise the likelihood. This is shown in Figure \ref{fig:contl} using an example light curve from our data set. As seen in the Figure the chosen hyper-parameters lie at the maximum log likelihood. Since the contours of variance and length scale only have one maximum (in the case of our light curves) we do not need to perform cross-validation to obtain the best hyper-parameters. The most computationally expensive part is inverting the \textit{covariance matrix} which requires a time $\mathcal{O}(N^3)$, and is the limiting factor for performing GP on large datasets.
	
	Once optimised, we can choose between different kernels by choosing the one with greater likelihood. 
	
	\subsection{Additional kernels}
\begin{figure}
		\centering
		\includegraphics[width=8cm]{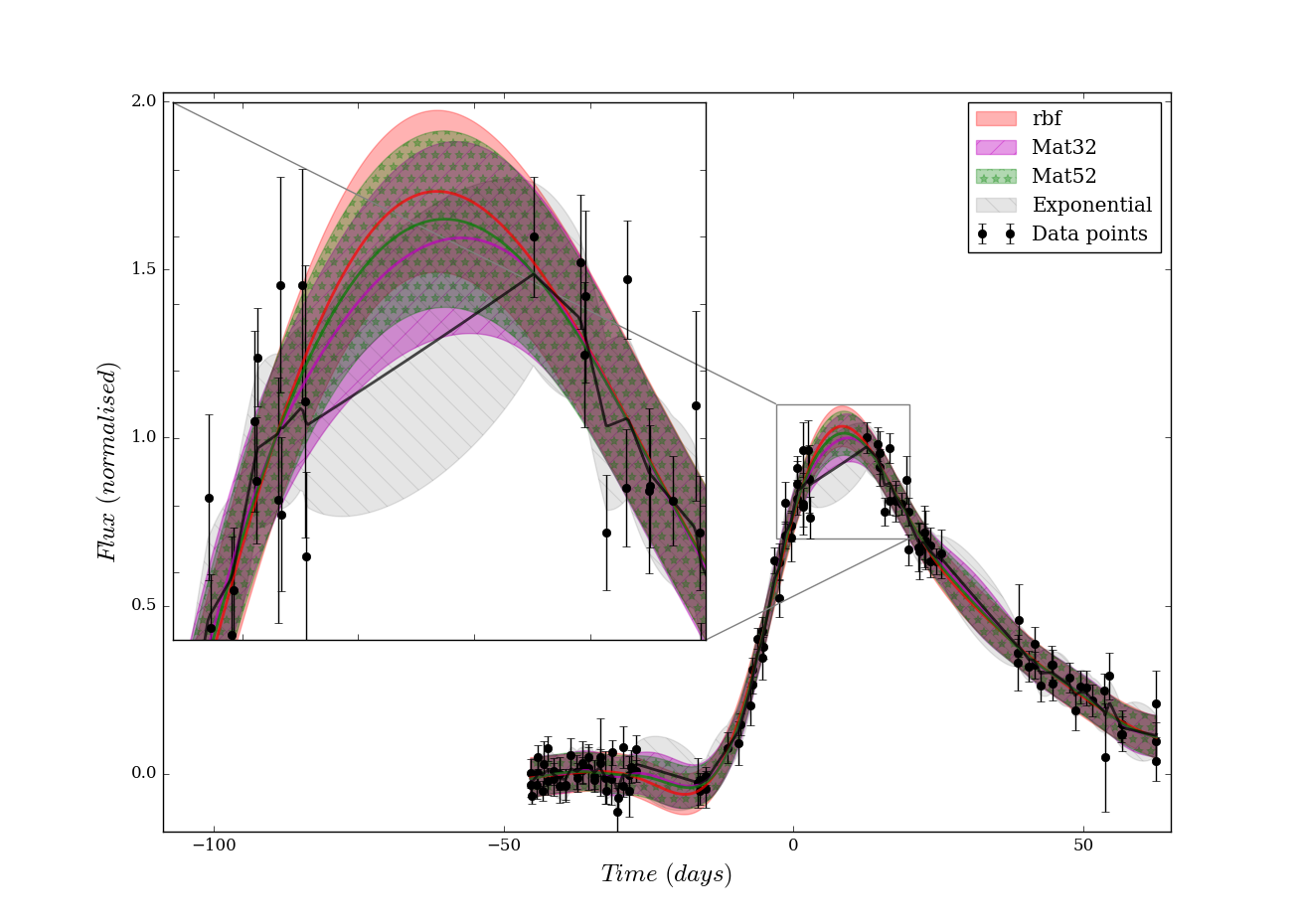}
		\caption{A plot of GP with different kernel for one of the supernovae light curves in our sample, iPTF13hpz. The solid lines (red, green, magenta and black for RBF, Mat\'ern with $\nu = 3/2$ and  $\nu =5/2$ and the exponential kernel respectively) show the latent functions and the shaded areas the $1 \sigma$ confidence interval. The zoomed in Section is the area near the maximum light where data points are missing.}\label{fig:kerneldiff}
	\end{figure}

	The square exponential kernel, shown in equation \ref{eq:rbf} forces the GP to be infinitely smooth, which may be unrealistic for some datasets. In our analysis we use the best kernel or a linear combination of different kernels used in the literature,
    to fit our data. The ones we use are introduced in the following Section. In Figure \eqref{fig:kerneldiff} we show an example of a light curve from our dataset under different kernels. Note that the biggest difference is when the data is sparse as shown in the inset-plot.
	
	\subsection{The Mat\'ern family}
	For a given $\nu$ the Mat\'ern kernels is defined as:
	\begin{equation}
		k_\nu({\bf{t}}, {\bf{t'}}) = \sigma^2 \left(\frac{1}{\Gamma(\nu)2^{\nu -1}}\left[\frac{\sqrt{2 \nu}}{l} |{\bf{t}}-{\bf{t'}}|\right]^{\nu} K_{\nu}\left(\frac{\sqrt{2 \nu}}{l}|{\bf{t}}-{\bf{t'}}|\right) \right),
	\end{equation}
	where $\nu$ regulates the smoothness of the function and is determined by how differentiable it is. $K_{\nu}$ is the modified Bessel function of second kind of order $\nu$, and $l$ is the characteristic length scale. From this family of kernels, we will select two cases of $\nu$:
	\begin{equation}
		k_{3/2}(t-t') = \sigma^2 \left(1 + \frac{\sqrt{3}|t-t'|}{l}\right) \exp \left(-\frac{\sqrt{3}|t-t'|}{l} \right) 
	\end{equation}
	and
	\begin{equation}
		k_{5/2}(t-t') = \sigma^2 \left(1 + \frac{\sqrt{5}|t-t'|}{l} + \frac{5(t-t')^2}{3l^2}\right) \exp \left(-\frac{\sqrt{5}|t-t'|}{l}. \right) 
	\end{equation}
	Note that as $\nu \rightarrow \infty$ we recover the squared exponential. 
	\subsection{Exponential kernel} 
	If $\nu = \frac{1}{2}$ we get the exponential kernel and the resulting function is continuous but non-differentiable. In 1D this corresponds to the Ornstein-Uhlenbeck process, a model of Brownian motion. It is defined as:
	\begin{equation}
		k(t-t') = \sigma^2 \exp \left(-\frac{|t-t'|}{l} \right).
	\end{equation}
	
	\subsection{White noise}
	To model a process that is not continuous we can use the white noise kernel, 
	\begin{equation}
		\mathscr{W} =  \sigma^2 \delta(t-t'),
	\end{equation}
	where $\delta$ is Dirac's $\delta$.
	
	\subsection{Linear kernel}
	Gaussian processes can also be reduced to a linear regression through the linear kernel, 
	\begin{equation}
		k(t, t') = \sigma^2 + t \cdot t'.
	\end{equation}
	\subsection{New kernels through linear combinations}
	We can also create new kernels by adding them in a linear combination. This can be used, for example, when both short scale and long scale phenomena are present in the data, which allows one kernel to have a different length-scale than the other. In particular we could use the linear kernel to capture the kernel to capture the global trend and the squared exponential to model local distortions. As mentioned in equation \eqref{eq:likely} there is a penalty for increased complexity preventing over-fitting.
    
	\subsection{Heteroscedastic Gaussian Processes}\label{Section:GPhetero}
	In its simplest formulation Gaussian Processes assumes that all the points have the same Gaussian noise and that this can be learned from the data. However in observational data we have known and different uncertainties. We can incorporate this by using a linear combination of kernels and a white noise kernel with $\sigma_i$ given by the uncertainties of the data:
	\begin{equation}
		\mathscr{W} =  \left[ \sum_i \sigma_i^2 \delta(t-t_i) \right] \delta(t-t'),
	\end{equation}
    
\section{Model comparison}\label{Section:Comparison}

In this paper we have used three different methods to fit the R-band light-curves of PTF and iPTF;  based on \citet{2017ApJ...838L...4Z}, SALT2 \citep{2007A&A...466...11G} and the GP template. Here we compare how statistically similar the light-curve fits are. With respect to the goodness of fit: 81 \%, 88 \% and 100\% of the fits for each method respectively have $\chi^2 / {\rm ndof} < 3$. The SALT2 fit values can be found Table \ref{table:salt2} and the values from \citet{2017ApJ...838L...4Z} in Table \ref{table:fitval}. We note that the $\chi^2$ is calculated only where each of the respective models are defined which is different for the 3 models (\citet{2017ApJ...838L...4Z} is only fitted to before the secondary maximum at $+15$ days, SALT2 \citep{2007A&A...466...11G} until $+50$ days and the GP template until $+80$ days), which makes a direct comparison harder.

\begin{figure}
		\includegraphics[width=10cm]{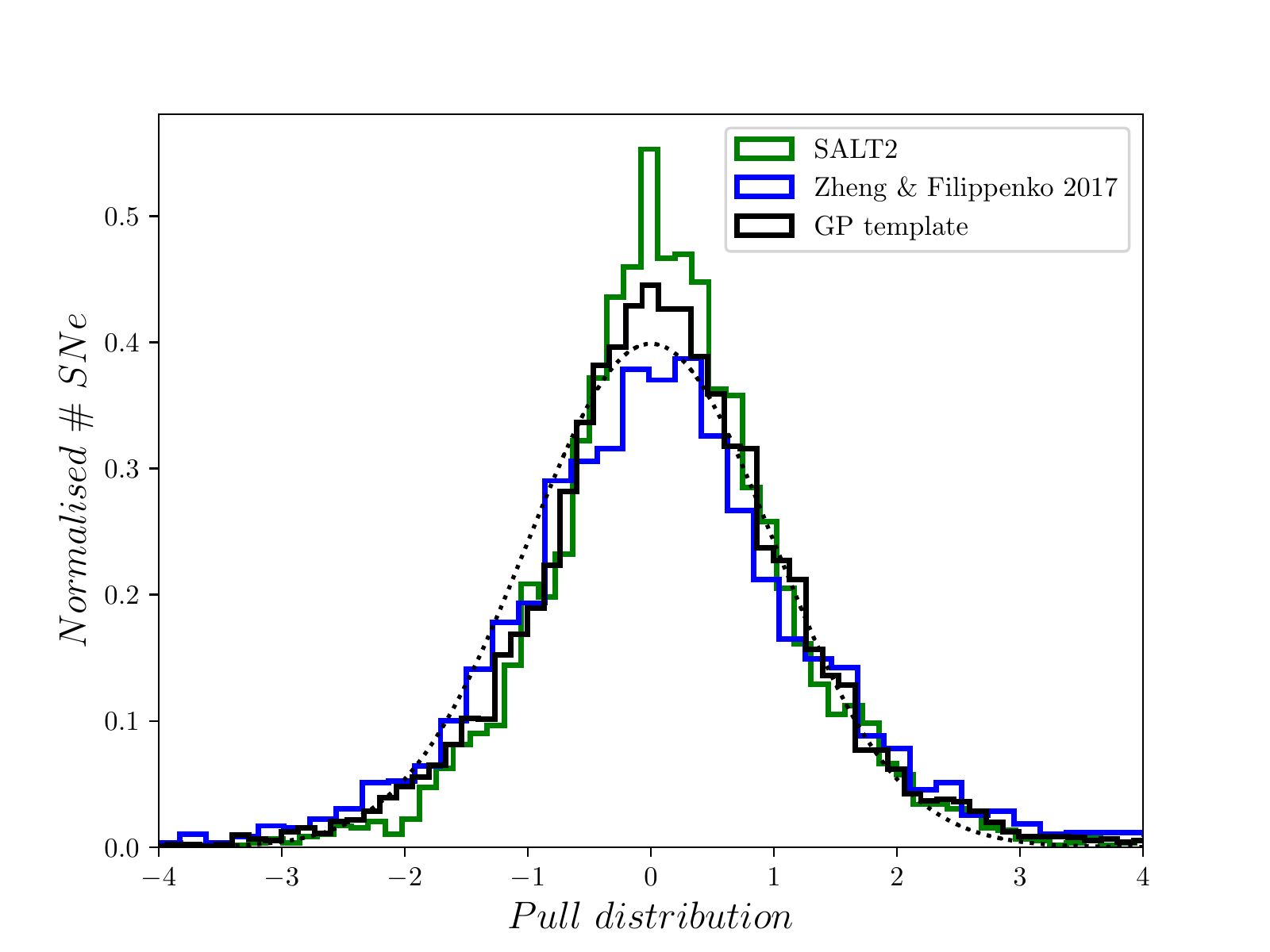}
		\caption{Here the Pull distributions, i.e. residuals/error is shown for the 3 different light-curve fitting methods used in this paper. The dotted line shows a Gaussian with $\sigma =1$, $\mu=0$ for comparison.}\label{fig:pull_comp}
	\end{figure}

To further investigate the $\chi^2 / {\rm ndof}$ differences between the models we compare the Pull distributions, i.e. the error weighted residuals. We first check whether they are consistent with a normal distribution for the three models, as seen in figure \ref{fig:pull_comp}. We also compare the Pull distributions as a function of phase and see no significant difference between the models. This is done in order to ascertain that none of the models fails to capture significant features of the light-curve within their entire model range, which could skew the distribution and would show up as a phase dependence. Since the models perform similarly we use the \citet{2017ApJ...838L...4Z} model because its parameters are easier to interpret physically. 

\onecolumn
\section{Tables of light-curve parameters}
\begin{longtable}{cccccc}
\caption{Best fit values for each supernova with their respective asymmetric errors.}\\
PTF name & $t_0$ (days) & A' & $\alpha_d$ & s & $\alpha_r$\\
\hline
\endhead
10aaju & $ -30.0 _{ -0.0 }^{+ 3.527 }$ & $ 7.183 _{ -1.762 }^{+ 1.209 }$ & $ -6.085 _{ -0.226 }^{+ 1.914 }$ & $ 0.529 _{ -0.028 }^{+ 0.185 } $& $ 2.695 _{ -0.138 }^{+ 0.11 }$ \\
\\
10aajv & $ -30.0 _{ -0.0 }^{+ 2.54 }$ & $ 31.952 _{ -4.062 }^{+ 4.753 }$ & $ -4.5 _{ -0.217 }^{+ 0.393 }$ & $ 0.3 _{ -0.0 }^{+ 0.008 } $& $ 3.176 _{ -0.058 }^{+ 0.061 }$ \\
\\
10aayx & $ -24.205 _{ -3.474 }^{+ 1.535 }$ & $ 1.349 _{ -0.043 }^{+ 0.524 }$ & $ -1.506 _{ -0.74 }^{+ 0.193 }$ & $ 3.0 _{ -1.305 }^{+ 0.0 } $& $ 1.573 _{ -0.06 }^{+ 0.212 }$ \\
\\
10abjm & $ -30.0 _{ -0.0 }^{+ 0.97 }$ & $ 9.614 _{ -1.175 }^{+ 1.471 }$ & $ -6.764 _{ -0.127 }^{+ 0.657 }$ & $ 0.452 _{ -0.021 }^{+ 0.041 } $& $ 2.926 _{ -0.087 }^{+ 0.095 }$ \\
\\
\end{longtable}

\twocolumn

\onecolumn
\begin{longtable}{ccccccccc}
\caption{Fit values and redshifts of the SNe in our sample.}\\
PTF name & Redshift & RA (J2000.0)& Dec (J2000.0) & Hubble residual (mag) & $m_R (peak)$ & $m_{Rerr} (mag)$ & Stretch & Stretch err\\
\hline
09alu & 0.071 & 214.573979 & 53.792253 & 0.28 & 18.88 & 0.025 & 1.28 & 0.016 \\
09dxo & 0.052 & 30.447754 & -7.093836 & 0.05 & 17.95 & 0.032 & 1.13 & 0.016 \\
10aaju & 0.078 & 161.553677 & 13.647916 & -0.13 & 18.68 & 0.007 & 1.03 & 0.012 \\
10aajv & 0.0638 & 147.18434 & 1.532675 & 1.29 & 19.64 & 0.01 & 0.74 & 0.015 \\
10aayx & 0.11 & 70.783009 & -4.726213 & 0.37 & 19.98 & 0.013 & 0.79 & 0.024 \\
10abjm & 0.04975 & 23.171613 & 34.635736 & 0.04 & 17.84 & 0.006 & 1.05 & 0.007 \\
\end{longtable}\label{table:fitvals}

\twocolumn
\onecolumn
\begin{longtable}{cccccccccc}
\caption{The SALT2 fit parameters for our SNe sample.}\\
Name & $z_{cmb}$ & t0 & $\sigma_{t0}$ & x0 & $\sigma_{x0}$ & x1 & $\sigma_{x1}$ & $\chi^2$ & ndof \\
\hline
09alu & 0.071 & 2455029.71 & 0.24 & 0.00 & 0.00 & -1.58 & 0.23 & 21.77 & 33.00 \\
09dxo & 0.052 & 2455093.07 & 1.15 & 0.00 & 0.00 & -0.27 & 1.33 & 8.13 & 11.00 \\
10aaju & 0.078 & 2455511.90 & 0.19 & 0.00 & 0.00 & 0.05 & 0.19 & 33.85 & 37.00 \\
10aajv & 0.0638 & 2455511.34 & 0.74 & 0.00 & 0.00 & 2.76 & 0.45 & 82.84 & 37.00 \\
10aayx & 0.11 & 2455516.02 & 0.82 & 0.00 & 0.00 & 1.02 & 0.91 & 65.75 & 27.00 \\
10abjm & 0.04975 & 2455533.89 & 0.14 & 0.00 & 0.00 & 0.22 & 0.14 & 25.58 & 39.00 \\
\end{longtable}
\twocolumn
\onecolumn
\begin{longtable}{ccc}
\caption{The Gaussian Processes template binned in bins of 1 day.}\\
\centering
Time (days) & $F/F_{peak}$ & $\sigma_F$ \\
\hline
-20.00 & 0.0063 & 0.0844 \\
-19.00 & 0.0148 & 0.0844 \\
-18.00 & 0.0293 & 0.0844 \\
-17.00 & 0.0514 & 0.0844 \\
-16.00 & 0.0825 & 0.0844 \\
-15.00 & 0.1237 & 0.0844 \\
\end{longtable}
\twocolumn
	
	\bsp	
	\label{lastpage}
\end{document}